\documentclass[aps,preprint,tightenlines,groupedaddress,nofootinbib,byrevtex,showpacs]{revtex4}
\usepackage{graphicx,comment}
\usepackage{epsfig}
\usepackage{amssymb,latexsym}
\usepackage{amsmath,amsbsy}

\begin{document}

\unitlength=1mm

\def\a{{\alpha}}
\def\b{{\beta}}
\def\d{{\delta}}
\def\D{{\Delta}}
\def\e{{\epsilon}}
\def\g{{\gamma}}
\def\G{{\Gamma}}
\def\k{{\kappa}}
\def\l{{\lambda}}
\def\L{{\Lambda}}
\def\m{{\mu}}
\def\n{{\nu}}
\def\o{{\omega}}
\def\O{{\Omega}}
\def\S{{\Sigma}}
\def\s{{\sigma}}
\def\th{{\theta}}

\def\ol#1{{\overline{#1}}}

\def\Dslash{D\hskip-0.65em /}
\def\dslash{{\partial\hskip-0.5em /}}
\def\vslash{{\rlap \slash v}}

\def\CPT{{$\chi$PT}}
\def\QCPT{{Q$\chi$PT}}
\def\PQCPT{{PQ$\chi$PT}}
\def\tr{\text{tr}}
\def\str{\text{str}}
\def\diag{\text{diag}}
\def\order{{\mathcal O}}
\def\vit{{\it v}}
\def\vD{\vit\cdot D}
\def\am{\alpha_M}
\def\bm{\beta_M}
\def\gm{\gamma_M}
\def\smb{\sigma_M}
\def\smt{\overline{\sigma}_M}
\def\tb{{\tilde b}}

\def\cS{{\mathcal S}}
\def\cC{{\mathcal C}}
\def\cB{{\mathcal B}}
\def\cT{{\mathcal T}}
\def\cQ{{\mathcal Q}}
\def\cL{{\mathcal L}}
\def\cO{{\mathcal O}}
\def\cA{{\mathcal A}}
\def\cH{{\mathcal H}}
\def\cF{{\mathcal F}}
\def\cG{{\mathcal G}}
\def\cE{{\mathcal E}}
\def\cJ{{\mathcal J}}
\def\cK{{\mathcal K}}
\def\cM{{\mathcal{M}_+}}

\def\Bbar{\overline{B}}
\def\Tbar{\overline{T}}
\def\cBbar{\overline{\cal B}}
\def\cTbar{\overline{\cal T}}
\def\cA{\mathcal A}
\def\pq{(PQ)}

\def\eqref#1{{(\ref{#1})}}

 
\title{Baryon Electromagnetic Properties in Partially Quenched Heavy Hadron Chiral Perturbation
  Theory}
\author{ Brian C.~Tiburzi}
\email[]{bctiburz@phy.duke.edu}
\affiliation{Department of Physics\\
Duke University\\
P.O.~Box 90305\\
Durham, NC 27708-0305}

\date{\today}

\begin{abstract} 
The electromagnetic properties of baryons containing a heavy quark 
are calculated at next-to-leading order in partially quenched 
heavy hadron chiral perturbation theory. Calculations 
are performed for three light flavors in the isospin limit
and additionally for two light non-degenerate flavors.
We use partially-quenched charge matrices that are easy to implement on the lattice.
The results presented are necessary for the light quark mass extrapolation 
and zero-momentum extrapolation of lattice QCD and partially quenched 
lattice QCD calculations of heavy hadron electromagnetic properties.
Additionally relations between the sextet electromagnetic form factors 
and transition form factors are derived.
\end{abstract}

\pacs{12.38.Gc}
\maketitle

\section{Introduction}

Calculating physical properties of hadrons from QCD is a formidable challenge. 
The strong interaction dynamics involved is highly non-perturbative and results 
in the confinement of quarks and gluons into color-neutral hadronic states. 
Electromagnetic probes have often been sought experimentally as a means to glimpse 
the charge and magnetism distributions of quarks within hadrons.
Electromagnetic observables, such as the charge radii and electromagnetic moments, provide
a physically intuitive glimpse at the structure of hadrons.
For baryons containing $b$ or $c$ quarks, however, the experimental difficulties in measuring
these observables are substantial.

The dynamics underlying singly heavy baryons involves a rich interplay of opposite limits of QCD:
heavy-quark physics and light-quark physics. 
Lattice QCD provides a first principles method for the calculation of QCD observables. 
These numerical simulations can determine the electromagnetic properties
of baryons containing a heavy quark and thereby provide a physical picture of these 
baryons in terms of their quark degrees of freedom. 
Calculation of singly heavy baryon observables (masses and weak transitions) have been pursued in quenched QCD 
(QQCD)~\cite{Alexandrou:1994dm,Bowler:1996ws,AliKhan:1999yb,Woloshyn:2000fe,Lewis:2001iz,Mathur:2002ce,Bowler:1997ej,Gottlieb:2003yb}.
Until now~\cite{Toussaint:2004cj}, partially quenched QCD (PQQCD) calculations have been very limited for any baryons.  
The appearance of quenched and partially quenched approximations to lattice QCD arises from numerically 
calculating the fermionic determinant. In QQCD, the time-costly numerical computation is radically simplified by replacing 
the determinant by a constant, whereas in PQQCD the determinant is calculated, but using larger quark masses 
for the quarks not coupled to external sources. In PQQCD, sea quark contributions are thus retained;
while in QQCD, the sea quarks have been effectively discarded. Due to restrictive computational time, 
the valence quark masses, too, cannot be set to their physical values and one must extrapolate down from the values 
used on the lattice.

To perform quark mass extrapolations, one must understand how QCD observables
behave as the quark masses vary. This can be done using chiral perturbation theory (\CPT), 
which is a model-independent effective theory for low-energy QCD.
This effective theory is written in terms of the pseudo-Goldstone bosons appearing from chiral 
symmetry breaking~\cite{Gasser:1983yg,Gasser:1985gg}. 
While these bosons are not massless, they remain light compared to the scale of chiral symmetry breaking
and dominate low-energy hadronic interactions.
The light quark masses appear as parameters of \CPT\ and enable
the determination of the quark mass dependence of QCD observables by matching onto the effective theory. 
Without external input (e.g.~experimental data or lattice QCD data), 
the effective theory has little predictive power because symmetries  
constrain the types of operators in the Lagrangian, but not the values of their coefficients.
While providing a guide to extrapolate lattice QCD data, \CPT\  
can also benefit in turn from the determination of its low-energy constants (LECs) from these data. 
Heavy quark symmetry furthermore can be combined with chiral symmetry to describe
hadrons formed from light quarks and a single heavy quark~\cite{Burdman:1992gh,Wise:1992hn,Yan:1992gz}.

For QQCD lattice simulations, quenched chiral perturbation theory (\QCPT) has been developed to aid in the 
extrapolation in valence quark masses~\cite{Morel:1987xk,Sharpe:1992ft,
Bernard:1992mk,Sharpe:1996qp,Labrenz:1996jy}. There is, however, no
general connection of QQCD observables to QCD because, for example, QQCD lacks an axial anomaly. 
Consequently \QCPT\ contains operators that do not have analogues in \CPT. 
Furthermore, the LECs of operators that have analogues
are numerically unrelated to their counterparts in \CPT. The sickness of the quenched approximation
can be treated by utilizing partially quenched lattice simulations. 
For such simulations, partially quenched chiral perturbation theory \PQCPT\ has been 
constructed to perform the extrapolation in both sea and valence quark 
masses~\cite{Bernard:1994sv,Sharpe:1997by,Golterman:1998st,Sharpe:2000bc,Sharpe:2001fh,Sharpe:2003vy}. 
Unlike QQCD, PQQCD retains an axial anomaly and the flavor singlet field can be integrated out. 
In further distinction to \QCPT, the LECs of \CPT\ appear in \PQCPT, and hence PQQCD lattice simulations
can be used to determine \CPT\ parameters. Much work has resulted from this possibility, e.g., in 
the baryonic sector 
see~\cite{Chen:2001yi,Beane:2002vq,Chen:2002bz,Arndt:2003vx,Arndt:2003ww,Arndt:2003we,Arndt:2003vd,Walker-Loud:2004hf,Tiburzi:2004rh}.

In this work we determine the electromagnetic properties 
of baryons containing a heavy quark at next-to leading order (NLO) in partially quenched heavy hadron \CPT. 
We determine the electromagnetic properties in three-flavor and two-flavor partially quenched simulations
and use the isospin limit in the former. Such expressions allow one to perform
the required extrapolation in light-quark masses. 
Additionally the electromagnetic LECs that appear in heavy hadron \CPT\ can be extracted at NLO from lattice data
and we explore how the arbitrariness of the light-quark charge matrix can be exploited to determine the LECs.
Furthermore we derive relations between the electromagnetic form factors and electromagnetic transition 
form factors of the sextet baryons. These relations follow from the decoupling of the heavy and light spin 
degrees of freedom.

The organization of the paper is as follows. First in Sec.~\ref{pqhhcpt}, we review the inclusion 
of baryons containing a heavy quark into \PQCPT. 
Next, the charge radii and magnetic moments of the $s_\ell = 0$ baryons are calculated in Sec.~\ref{triplet}. 
The charge radii, magnetic moments and electric quadrupole moments of the $s_\ell = 1$ baryons are calculated 
in Sec.~\ref{six}. The corresponding results for $SU(4|2)$ with non-degenerate light quarks 
are presented in Appendix~\ref{pqsutwo}. For reference, in Appendix~\ref{cpt} we give 
the electromagnetic properties of these baryons in $SU(3)$ and $SU(2)$ \CPT.  
In Appendix~\ref{s:HQ}, we use heavy quark symmetry to derive relations between the sextet baryon form factors; and 
in Appendix~\ref{s:charge}, we present results for and comment on the commonly used form of the partially-quenched charge matrix. 
The summary, Sec.~\ref{summy}, highlights the goal of understanding the electromagnetic properties of 
the triplet and sextet baryons from lattice QCD, and contrasts the situation with that of baryons 
formed from three light quarks.

\section{\PQCPT\ for heavy hadrons} \label{pqhhcpt}

In this section, we present the formulation of \PQCPT\ for heavy hadrons in $SU(6|3)$. 
We begin by reviewing the set up of PQQCD. Next we describe the pseudo-Goldstone
mesons of \PQCPT, and finally include the heavy hadrons
into this partially quenched theory.

\subsection{PQQCD} \label{subpqqcd}

The light-quark sector Lagrangian in PQQCD is given by
\begin{equation}
\mathfrak{L} = \sum_{j,k=1}^9 \ol{q}_j \left(
  i\Dslash - m_q \right)_{jk} q_k
.\label{eq:pqqcdlag}
\end{equation}
This differs from the QCD Lagrangian by the addition
of six extra quarks; three bosonic ghost quarks, ($\tilde u,
\tilde d, \tilde s $), and three fermionic sea quarks, ($j, l, r$), in addition
to the light physical quarks ($u, d, s$).  The nine quark fields transform in the 
fundamental representation of the graded group $SU(6|3)$~\cite{BahaBalantekin:1980pp,BahaBalantekin:1981qy}.  
They appear in the nine-component vector
\begin{equation}
q = (u, d, s, j, l, r, \tilde{u}, \tilde{d}, \tilde{s})^{\text{T}}
.\end{equation}
These quark fields obey graded equal-time commutation relations
\begin{equation}
q^\a_i(\mathbf x) q^{\beta \dagger}_j(\mathbf y) -
(-1)^{\eta_i \eta_j} q^{\b \dagger}_j(\mathbf y) q^\a_i(\mathbf x) =
\d^{\a \b} \d_{ij} \d^3 (\mathbf {x-y}),
\label{eq:qetcr}\end{equation}
where $\a, \beta$ and $i,j$ are spin and flavor indices, respectively.
The graded equal-time commutation relations which vanish can be written similarly.
The different statistics of the PQQCD quark fields are reflecting in grading factors $\eta_k$
employed above, where
\begin{equation}
   \eta_k
   = \left\{ 
       \begin{array}{cl}
         1 & \text{for } k=1,2,3,4,5,6 \\
         0 & \text{for } k=7,8,9
       \end{array} 
     \right.
.\end{equation}
In the isospin limit $m_u = m_d$, the quark mass matrix of $SU(6|3)$ is given by
\begin{equation}
m_q = \diag(m_u, m_u, m_s, m_j, m_j, m_r, m_u, m_u, m_s).
\end{equation}
Notice that each valence quark mass is degenerate with the corresponding ghost quark's mass. 
This equality maintains an exact cancellation in the path integral between their
respective determinants. Putting only valence quarks in the external states, the contributions to observables from 
disconnected quark loops come from the sea sector of PQQCD. Thus one can separate valence 
and sea contributions as their names suggest, and independently vary the masses of the valence and sea sectors. 

The light quark electric charge matrix $\cQ$ is not uniquely defined in PQQCD~\cite{Golterman:2001qj}.
By imposing the charge matrix $\cQ$ to be supertraceless in $SU(6|3)$, no new operators involving the singlet component
are introduced. This can be accomplished with the choice 
\begin{equation} 
\cQ = \diag ( q_u, q_d, q_s, q_j, q_l , q_r, q_u, q_d, q_s ), 
\end{equation}
where to maintain supertracelessness $q_j + q_l + q_r = 0$. 
Notice that when the sea quark masses are made degenerate with 
the valence quark masses, QCD is only recovered for the specific choice 
$q_u = q_j = \frac{2}{3}$, and $q_d = q_s = q_l = q_r + - \frac{1}{3}$. 
As we shall detail below, one can use unphysical charges for both valence and sea quarks 
as a means to determine the LECs. 
The results and problems for the commonly used partially-quenched form of $\cQ$ proposed in~\cite{Chen:2001yi} 
are presented in Appendix~\ref{s:charge}.

\subsection{Mesons of \PQCPT} \label{mesonpqcpt}

For massless light quarks, 
the theory described by the Lagrangian in
Eq.~(\ref{eq:pqqcdlag}) has the symmetry
$SU(6|3)_L \otimes SU(6|3)_R \otimes U(1)_V$, that is
assumed to be spontaneously broken to 
$SU(6|3)_V \otimes U(1)_V$ in analogy with QCD.  
One can build an effective low-energy theory of PQQCD by
perturbing about the physical vacuum state. 
This theory is PQ$\chi$PT, and the dynamics of the pseudo-Goldstone mesons appearing from chiral 
symmetry breaking are described at leading order in the chiral expansion by the Lagrangian
\begin{equation}
  \mathfrak{L} =
    \frac{f^2}{8} \str \left(
      D^\mu \S^\dagger D_\mu \S \right)
      + \l  \, \str \left( m_q \S^\dagger + m_q^\dagger \S \right)
,\label{eq:pqbosons}
\end{equation}
where the field 
\begin{equation}
  \S = \exp \left( \frac{2 i \Phi}{f} \right) = \xi^2,
\end{equation}
and the meson fields appear in the $SU(6|3)$ matrix
\begin{equation}
    \Phi =
    \begin{pmatrix}
      M & \chi^\dagger\\
      \chi & \tilde M\\
    \end{pmatrix}. \label{eq:mesonmatrix}
\end{equation}
The $M$ and $\tilde M$ matrices contain bosonic mesons (with quantum numbers of $q \bar{q}$ pairs and 
$\tilde{q} \bar{\tilde{q}}$ pairs, respectively), while the $\chi$ and $\chi^\dagger$
matrices contain fermionic mesons (with quantum numbers of $\tilde q \bar{q}$
pairs and $q \bar{\tilde{q}}$ pairs, respectively).
The upper $3 \times 3$ block of the matrix $M$ contains the familiar
pions, kaons, and eta, while the remaining components consist of mesons formed
from one or two sea quarks, see e.g.~\cite{Chen:2001yi}.
The operation $\str()$ in Eq.~\eqref{eq:pqbosons} is a graded flavor trace.
The gauge covariant derivative is defined by $D_\mu \Sigma = \partial_\mu \Sigma + i e \mathcal{A}_\mu [ \cQ, \Sigma]$,
where $\mathcal{A}_\mu$ is the photon field.
To leading order, one finds that mesons with quark content $qq'$ are canonically normalized
when their masses are given by
\begin{equation}
m^2_{qq'} = \frac{4 \lambda}{f^2} (m_q + m_{q'}) 
\label{eq:pqmesonmass}.
\end{equation}

The flavor singlet field is
$\Phi_0 = {\rm str}( \Phi ) / {\sqrt 6}$, and because
PQQCD has a strong axial anomaly $U(1)_A$, 
the mass of the singlet field has been taken to be
on the order of the chiral symmetry breaking scale, and thus $\Phi_0$ 
has been integrated out of Eq.~\eqref{eq:pqbosons}. 
The resulting $\eta$ two-point correlation functions, however, 
deviate from their familiar form in \CPT. 
In calculating the  electromagnetic properties of heavy hadrons, the results do not explicitly
depend on the form of the flavor-neutral propagator.

\subsection{Baryons containing heavy quarks in \PQCPT} \label{hqbpqcpt}

Heavy quark effective theory (HQET) and \PQCPT\ can be combined to describe
the interactions of heavy hadrons and pseudo-Goldstone mesons. 
Let $m_Q$ denote the mass of the heavy quark, where $Q = c$ or $b$. 
In the limit $m_Q \to \infty$, baryons containing a heavy quark and two light quarks 
(and indeed all heavy hadrons) are classified by the spin of their light degrees of freedom, $s_\ell$, 
because the heavy quark's spin decouples from the system.
The inclusion of baryons containing a heavy quark into \CPT\ was carried out 
in~\cite{Yan:1992gz,Cho:1992gg,Cho:1992cf} and their magnetic moments were calculated to one-loop 
order in~\cite{Savage:1994zw,Banuls:1999mu}. The quenched chiral theory for baryons with a heavy quark
was written down in~\cite{Chiladze:1997uq} and the partially quenched theory 
was investigated in~\cite{Arndt:2003vx,Tiburzi:2004kd}.

Consider first the baryons with $s_\ell = 0$. 
To include these spin-$\frac{1}{2}$ baryons into \PQCPT, 
we use the method of interpolating fields to classify their representations of 
$SU(6|3)$~\cite{Labrenz:1996jy,Savage:2001dy,Chen:2001yi,Beane:2002vq,Arndt:2003vx}. 
The $s_\ell = 0$ baryons are described by the field\footnote{%
As we largely work to leading order in $1/m_Q$, the label $Q$ is omitted from all baryon tensors.
}
\begin{equation}
\cT^\gamma_{ij} 
\sim 
Q^{\gamma,c} 
\left( 
q_i^{\a,a} q_j^{\b,b}
+ 
q_i^{\b,b} q_j^{\a,a} 
\right)
\varepsilon_{abc} (C\gamma_5)_{\a\b}
,\label{eq:Tinterp}
\end{equation}
where $i$ and $j$ are flavor indices; $a$, $b$, and $c$ are color indices; and $\a$, $\b$, and $\gamma$ 
are spin indices. The tensor $\cT$ has the symmetry property 
\begin{equation}
\cT_{ij} = (-)^{\eta_i \eta_j} \cT_{ji}
\end{equation}
and forms a $\mathbf{39}$-dimensional representation of $SU(6|3)$. 
These states are conveniently classified under the quark sectors acted upon, 
thus we use the super-algebra terminology of~\cite{Hurni:1981ki,Chen:2001yi}. 
Under $SU(3)_{\text{val}} \otimes SU(3)_{\text{sea}} \otimes SU(3)_{\text{ghost}}$, 
the ground floor, level A transforms as a $(\mathbf{3},\mathbf{1},\mathbf{1})$ and 
contains the familiar $s_\ell = 0$ QCD baryon tensor $T_{ij}$, 
i.e.~$\cT_{ij} = T_{ij}$, when the indices are restricted to the range $1-3$. 
With our conventions, we have 
\begin{equation}
T_{ij} 
= 
\frac{1}{\sqrt{2}}
\begin{pmatrix}
0              &     \L_Q      & \Xi_Q^{+\frac{1}{2}} \\
- \L_Q         &        0      & \Xi_Q^{-\frac{1}{2}} \\
- \Xi_Q^{+\frac{1}{2}} & -\Xi_Q^{-\frac{1}{2}} &      0     
\end{pmatrix}_{ij}
\label{eq:TSU3},\end{equation}
and the superscript labels the $3$-projection of isospin. 
The first floor of level A contains nine baryons that transform 
as a $(\mathbf{3},\mathbf{1},\mathbf{3})$, while the ground floor of
level B contains nine baryons that transform as a 
$(\mathbf{3},\mathbf{3},\mathbf{1})$. The remaining floors and levels
are not necessary for our calculation.

Next we consider the $s_\ell = 1$ baryons. These spin $\frac{1}{2}$ and $\frac{3}{2}$ baryons
are degenerate in the heavy quark limit and can be described by one field $\cS^\mu_{ij}$.  
The interpolating field for these baryons is
\begin{equation}
\cS^{\mu, \gamma}_{ij} 
\sim 
Q^{\gamma,c} 
\left( 
q_i^{\a,a} q_j^{\b,b}
- 
q_i^{\b,b} q_j^{\a,a} 
\right)
\varepsilon_{abc} (C\gamma^\mu)_{\a\b}
,\label{eq:Sinterp} 
\end{equation}
where the tensor satisfies the symmetry
\begin{equation}
\cS^\mu_{ij} = (-)^{1 + \eta_i \eta_j } \cS^\mu_{ji}
,\end{equation}
and makes up a $\mathbf{42}$-dimensional representation of $SU(6|3)$. 
Under $SU(3)_{\text{val}} \otimes SU(3)_{\text{sea}} \otimes SU(3)_{\text{ghost}}$, 
the ground floor, level A transforms as a $(\mathbf{6},\mathbf{1},\mathbf{1})$ and 
contains the familiar $s_\ell = 1$ QCD baryon tensor $S^\mu_{ij}$, 
i.e.~$\cS^\mu_{ij} = S^\mu_{ij}$, when the indices are restricted to the range $1-3$.
The QCD flavor tensor is
\begin{equation}
S^\mu_{ij} = \frac{1}{\sqrt{3}} (v^\mu + \gamma^\mu) \gamma_5 \, B_{ij}  + B^{* \mu}_{ij}
,\label{eq:Sdecomp}
\end{equation} 
with
\begin{equation}
B_{ij} 
= 
\begin{pmatrix}
\Sigma_Q^{+1}                  &  \frac{1}{\sqrt{2}} \Sigma_Q^0   & \frac{1}{\sqrt{2}} \Xi_Q^{\prime + \frac{1}{2}} \\
\frac{1}{\sqrt{2}} \Sigma_Q^0  &  \Sigma_Q^{-1}                   & \frac{1}{\sqrt{2}} \Xi_Q^{\prime - \frac{1}{2}} \\
\frac{1}{\sqrt{2}} \Xi_Q^{\prime + \frac{1}{2}}           &  \frac{1}{\sqrt{2}} \Xi_Q^{\prime - \frac{1}{2}}            & \Omega_Q
\end{pmatrix}_{ij}
\label{eq:BSU3},\end{equation}
and similarly for the $B^{*\mu}_{ij}$. Above $v^\mu$ is the velocity vector of the heavy hadron, 
and we suppress velocity labels on all heavy hadron fields throughout. 
Again the superscript on these states labels the $3$-projection of isospin. 
The first floor of level A contains nine baryons that transform 
as a $(\mathbf{3},\mathbf{1},\mathbf{3})$, while the ground floor of
level B contains nine baryons that transform as a 
$(\mathbf{3},\mathbf{3},\mathbf{1})$. The remaining floors and levels
are not necessary for our calculation.

The free Lagrangian for the $\cT$ and $\cS^\mu$ fields is given by
\begin{eqnarray}
\mathfrak{L} 
&=& -
i \left( \ol \cS {}^\mu v \cdot D \cS_\mu \right)
+ 
\D \left( \ol \cS {}^\mu \cS_\mu \right)
+ 
\l_1 \left( \cS {}^\mu  \cM \cS_\mu \right)
+ 
\l_2 \left( \cS {}^\mu \cS_\mu \right) \str \cM
\notag \\
&& + 
i \left( \ol \cT v \cdot D \cT \right) 
+ 
\l_3 \left( \ol \cT \cM \cT \right)
+ 
\l_4 \left( \ol \cT \cT \right) \str \cM 
\label{eq:STfree}
.\end{eqnarray}
We have employed $()$-notation for flavor contractions that are invariant under
chiral transformations. The relevant contractions can be found in~\cite{Arndt:2003vx}.
The mass of the $\cT$ field has been absorbed into the static phase of the heavy hadron fields. 
Thus the leading-order mass splitting $\D$ appears as the mass of the $\cS^\mu$. This splitting remains finite as 
$\{ m_Q \to \infty, m_q \to 0 \}$ and cannot be removed by field redefinitions due to the interaction of the 
$\cT$ and $\cS^\mu$ fields. We treat $\D \sim m_\pi$ in our power counting. 
The Lagrangian contains the chiral-covariant derivative $D^\mu$, 
the action of which is identical on $\cT$ and $\cS^\mu$ fields and has the form~\cite{Arndt:2003vx}
\begin{equation}
\left( D^\mu \cT \right)_{ij}
=
\partial^\mu \cT_{ij} 
+ 
V^\mu_{ii'} \cT_{i'j} 
+ 
(-)^{\eta_i (\eta_j + \eta_{j'})}
V^\mu_{jj'} \cT_{ij'} \label{eq:covD}
.\end{equation}
The vector and axial-vector meson fields are defined by
\begin{equation}
V^\mu = \frac{1}{2} \left( \xi \partial^\mu \xi^\dagger + \xi^\dagger \partial^\mu \xi \right), 
\qquad 
A^\mu = \frac{i}{2} \left( \xi \partial^\mu \xi^\dagger - \xi^\dagger \partial^\mu \xi \right)
.\end{equation} 
The mass operator has the usual definition
\begin{equation}
\cM = \frac{1}{2} \left( \xi m_q \xi + \xi^\dagger m_q \xi^\dagger \right)
.\end{equation}
Adding electromagnetic interactions to the covariant derivative can be accomplished by the replacement
$V^\mu_{ij} \to V^\mu_{ij} + i e \mathcal{A}^\mu ( \cQ_Q \delta_{ij} / 2 + \cQ_{ij})$ in Eq.~\eqref{eq:covD}, where $\cQ_Q$ is 
the heavy quark charge and the factor of one-half normalizes the action of the Kronecker delta.
Gauging of the axial-vector fields is not necessary to the order we work.

The Lagrangian that describes the interactions of the $\cT$ and $\cS^\mu$ fields with 
the pseudo-Goldstone modes is given at leading order by the Lagrangian
\begin{equation}
\mathfrak{L} 
= 
i g_2 
\left( 
\ol \cS {}^\mu v^\nu A^\rho \cS^\sigma 
\right)
\varepsilon_{\mu \nu \rho \sigma}
+
\sqrt{2} \, g_3 
\left[ 
\left( 
\ol \cT A^\mu \cS_\mu 
\right) 
+ 
\left(
\ol \cS {}^\mu A_\mu \cT
\right)
\right]
\label{eq:STM}
.\end{equation}
The LECs appearing above, the $\l_j$, $g_2$, and $g_3$, all have 
the same numerical values as those used in $SU(3)$ heavy hadron \CPT. 
This is described in Appendix \ref{cpt}.

\section{Electromagnetic properties  of the $s_{\ell} = 0$ baryons} \label{triplet}

In this section, we calculate the charge radii of the $s_\ell = 0$ baryons 
in \PQCPT\ to leading order in the heavy quark expansion. Additionally the magnetic moments
are calculated to next-to-leading order in the heavy quark expansion in order to 
ascertain the leading light-quark mass dependence.  
Baryon matrix elements of the electromagnetic current $J^\mu$ can be parametrized in terms 
of two form factors $F_1$ and $F_2$. In the heavy hadron formalism, such a decomposition is
\begin{equation} \label{eq:Tdecomp}
\langle \ol T(p') | J^\mu | T(p) \rangle 
= 
\ol u (p') 
\left[ 
v^\mu F_1(q^2)
+
\frac{i \sigma^{\mu \nu} q_\nu}{2 M_T} F_2(q^2)
\right]
u(p)
,\end{equation}
with $q = p' - p$ as the momentum transfer. The Dirac form factor $F_1(q^2)$ is normalized
at zero momentum transfer to the baryon charge $Q_T$ in units of the electron charge $e$, namely
$F_1(0) = Q_T$. The baryon charge is a sum of the heavy quark charge $\cQ_Q$, and the total charge of 
the light degrees of freedom $\cQ_T$, i.e.~$Q_T = \cQ_Q + \cQ_T$. Keep in mind that the light-quark
electric charge matrix leaves the valence quarks with arbitrary charges, hence $Q_T$ and $\cQ_T$ need not
have their physical values.
Ignoring the Dirac contribution, the magnetic moment is given by
\begin{equation}
\mu = F_2(0)
,\end{equation}
and the Dirac charge radius by
\begin{equation}
\langle r^2 \rangle  = 6 \frac{d}{dq^2} F_1(q^2) \Bigg|_{q^2=0}
.\end{equation}

These electromagnetic observables receive two types 
of contributions: short range contributions, which arise from local operators in the effective theory; 
and long range contributions, which arise from loop graphs involving the pseudo-Goldstone modes. 
The local operators characterizing the short distance electromagnetic interactions of 
heavy quark baryons come in two forms. First there are electromagnetic operators corresponding
to the current interacting with the heavy quark. These operators can be found by matching the HQET operators 
onto the chiral theory \cite{Cho:1992nt,Cheng:1992xi} and thus have fixed coefficients. 
Secondly there are operators which characterize the short distance interaction of the current with the light degrees of freedom. 
Chiral symmetry dictates only the form of these operators, leaving them with undetermined LECs.

The HQET part of the short distance dynamics comes from the anomalous magnetic moment and charge radius of the heavy quark. 
In \PQCPT\ the magnetic moment operator is matched in the $s_\ell = 0$ baryon sector onto the term~\cite{Cho:1992nt,Cheng:1992xi}\footnote{%
We use $F^{\mu \nu} = \partial^\mu \mathcal{A}^\nu - \partial^\nu \mathcal{A}^\mu$ for the electromagnetic field-strength tensor. 
}
\begin{equation} \label{eq:hqmmT}
\mathfrak{L} = - \frac{e \mu_Q}{4 m_Q} \left( \ol \cT \sigma_{\mu \nu} \cT \right) F^{\mu \nu} 
,\end{equation}
where the heavy quark magnetic moment $\mu_Q$ consists of the Dirac piece plus the anomalous part, 
i.e.~$\mu_Q = \cQ_Q + \frac{2 \a_s(m_Q)}{3 \pi} + \ldots$.
The charge radius contribution from the photon coupling to the heavy quark arises from the electromagnetic 
Darwin term~\cite{Caswell:1985ui,Manohar:1997qy} that appears in the HQET Lagrangian as 
\begin{equation}
\mathfrak{L} = - \frac{e \cQ_Q}{8 m_Q^2} \, \ol Q \, v_\mu D_\nu F^{\mu \nu} Q
.\end{equation}
In the $s_\ell = 0$ baryon sector, this HQET operator matches onto the \PQCPT\ term
\begin{equation}
\mathfrak{L} = - \frac{e \cQ_Q}{8 m_Q^2} \left( \ol \cT \cT \right)  v_\mu \partial_\nu F^{\mu \nu}
,\end{equation}
which contributes to the baryon charge radii.

Now we assess short-distance contributions to electromagnetic observables from the light degrees of freedom. 
In \PQCPT\ the magnetic moment contribution from the light quarks is suppressed by $\L_{QCD} / m_Q$ because
the light degrees of freedom have $s_\ell = 0$. Thus the leading magnetic moment operator has the form
\begin{equation} \label{eq:mmT}
\mathfrak{L} = - \frac{e \mu_T \L_{QCD}}{4 \L_\chi m_Q} \left( \ol \cT \cQ \sigma_{\mu \nu} \cT \right) F^{\mu \nu}
.\end{equation}
The local contributions in \PQCPT\ to the charge radii of the $s_\ell = 0$ baryons arise from the dimension-six 
operator that is contained in the Lagrangian
\begin{equation} \label{eq:chT}
\mathfrak{L} = - \frac{e c_T }{\L_\chi^2} \left( \ol \cT \cQ \cT \right) v_\mu \partial_\nu F^{\mu \nu}
.\end{equation}
Although the quark charges are arbitrary,
the LECs  $\mu_T$ and $c_T$ of these two $SU(6|3)$ \PQCPT\ operators have the same numerical values as the 
LECs in three flavor \CPT. We make this connection in Appendix~\ref{cpt}.

\begin{figure}
\epsfig{file=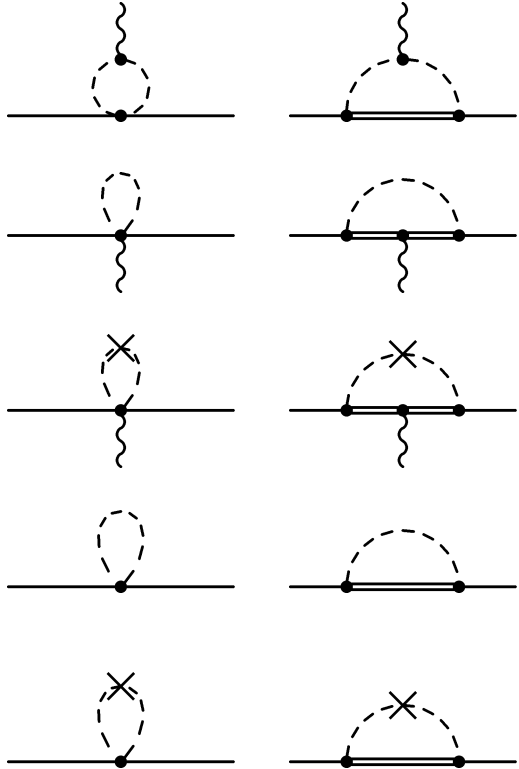}
\caption{Loop diagrams contributing to the charge radii of the $s_\ell = 0$ baryons.  Mesons are denoted 
by a dashed line, flavor neutrals (hairpins) by a crossed dashed line, and the photon by a wiggly line.  A thin (thick) 
solid line denotes an $s_\ell = 0$ ($ s_\ell = 1$) baryon.  The diagrams in the first row contribute to the Dirac form factor. 
The remaining diagrams with a photon have no $q^2$-dependence. These, along with the
wavefunction renormalization diagrams in the last two rows, ensure non-renormalization of the electric charge.}
\label{F:Tloops}
\end{figure}

The long distance contributions to these electromagnetic observables arise from the one-loop diagrams 
depicted in Fig.~\ref{F:Tloops}. The diagrams involve the pseudo-Goldstone mesons and the vertices
are generated from the leading-order interaction Lagrangian Eq.~\eqref{eq:STM}. 
These loops, however, do not contribute to the magnetic moments of the $s_\ell = 0$ baryons.
The leading light-quark mass dependence of the magnetic moments enters from heavy quark symmetry breaking
terms. We shall address these terms below. 
 By contrast, the charge radii receive contributions from the one-loop diagrams depicted in the figure. 
These diagrams are divergent and the resultant scale dependence is absorbed by a counterterm
of the form in Eq.~\eqref{eq:chT}. 
In this calculation and throughout this work, 
we use dimensional regularization with a modified minimal subtraction scheme, 
where we have only subtracted terms proportional to 
\begin{equation}
\frac{1}{\varepsilon} - \gamma_E + 1 + \log 4 \pi
.\end{equation}

To compactly write out the $q^2$-dependence of the form factors,  we define the abbreviation
\begin{equation}
P_\phi = \sqrt{1 - x (1 -x) \frac{q^2}{m_\phi^2}}
.\end{equation}
Combining the local and loop contributions, 
for the Dirac form factor we find
\begin{eqnarray}
F_1(q^2) 
&=&
Q_T + 
\left( 
\cQ_Q \frac{1}{8 m_Q^2} + \cQ_T \frac{c_T}{\L_\chi^2} 
\right)
q^2 
\notag \\
&& +
\frac{1}{8 \pi^2 f^2} 
\sum_\phi \a_\phi^T 
\int_0^1 dx
\left[ 
-\frac{q^2}{6} \log \frac{m_\phi^2}{\mu^2} 
+
2 m_\phi^2 P_\phi^2 \log P_\phi
\right]
\notag \\
&& + 
\frac{3 g_3^2}{(4 \pi f)^2} \sum_\phi \a_\phi^T
\int_0^1 dx 
\Bigg[
\cJ(m_\phi P_\phi, \D, \mu) - \cJ(m_\phi, \D, \mu)
- \frac{2}{9} q^2 
\notag \\
&& \phantom{morespacesmorespaces}
- \frac{2}{3} x (1-x) q^2 \,
\cG (m_\phi P_\phi, \D, \mu) 
\Bigg] \label{eq:F1T}
.\end{eqnarray}
The non-analytic functions arising from loop integrals have been abbreviated in the above expression by
\begin{equation}
\cG(m, \d, \mu) 
= 
\log \frac{m^2}{\mu^2} 
- 
\frac{\d}{\sqrt{\d^2 - m^2}} 
\log \frac{\d - \sqrt{\d^2 - m^2 + i \varepsilon}}{\d + \sqrt{\d^2 - m^2 + i \varepsilon}}
\end{equation}
and
\begin{equation}
\cJ(m, \d, \mu) 
= 
(m^2 - 2 \d^2) 
\log \frac{m^2}{\mu^2} 
+
2 \d \sqrt{\d^2 - m^2}
\log \frac{\d - \sqrt{\d^2 - m^2 + i \varepsilon}}{\d + \sqrt{\d^2 - m^2 + i \varepsilon}}
.\end{equation}
The sums in Eq.~\eqref{eq:F1T} are over loop mesons $\phi$ of mass $m_\phi$. The coefficients
$\a_\phi^T$ appear in Table~\ref{t:TPQQCD-A} and depend on the particular $s_\ell = 0$ baryon state $T$. 
Notice the charges $q_j$ and $q_l$ do not explicitly appear in these coefficients.  
In the isospin limit, they always enter in the combination $q_j + q_l$ which is the same as $-q_r$ by supertracelessness of $\cQ$.

From the table, we see that with efficacious choices for the charges of valence and sea quarks contributions from 
certain loop mesons can be eliminated thereby simplifying the chiral extrapolation. 
For the $\L_Q$ baryon, we can eliminate all but one loop meson; while for the $\Xi_Q$ baryons,
we can reduce the competition from four loop mesons down to two. 
Alternately there is a choice of sea charges which decreases computation time. 
One can preserve the supertracelessness of the charge matrix with the choice $q_j = q_l = q_r = 0$. 
For this choice, the lattice practitioner does not calculate closed quark loops with current insertion. 
Furthermore either of the valence charges can be taken to zero, and the LECS can still be extracted from the 
resulting electromagnetic properties. Once the LECs are known, physical predictions can be made.

\begin{table}
\caption{The coefficients $\a^T_\phi$ in $SU(6|3)$ \PQCPT. 
Coefficients are listed for the $s_\ell = 0$ baryon states $T$, and are grouped into contributions from loop mesons
with mass $m_\phi$.}
\begin{tabular}{l | c c c c }
    & $ \qquad ju \qquad \;$ & $ \qquad ru \qquad \;$ 
    & $\qquad js \qquad \;$  & $\qquad rs \qquad \;$ \\
\hline
$\L_Q$     
	   &  $q_u + q_d + q_r$ & $\frac{1}{2}(q_u + q_d) - q_r$  
           &  $0$ & $0$ \\
$\Xi^{+ \frac{1}{2}}_Q$     
	   &  $q_u  + \frac{1}{2} q_r$ & $\frac{1}{2} (q_u - q_r)$  
           &  $q_s + \frac{1}{2} q_r$ & $\frac{1}{2} (q_s - q_r)$ \\
$\Xi^{- \frac{1}{2}}_Q$     
           &  $q_d + \frac{1}{2} q_r$ & $\frac{1}{2} (q_d - q_r )$  
           &  $q_s + \frac{1}{2} q_r$ & $\frac{1}{2} (q_s - q_r )$ \\
\end{tabular}
\label{t:TPQQCD-A}
\end{table}

\begin{figure}
\epsfig{file=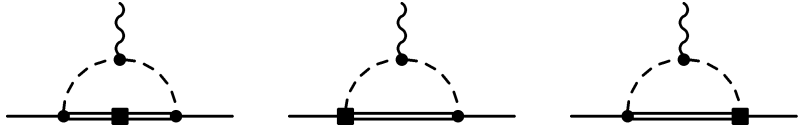}
\caption{Loop diagrams contributing to the leading light-quark mass dependence of the $s_\ell = 0$ baryons' magnetic moments.  
Mesons are denoted by a dashed line, the photon by a wiggly line, and a thin (thick) 
solid line denotes an $s_\ell = 0$ ($ s_\ell = 1$) baryon.  Heavy quark symmetry breaking vertices are depicted 
by squares and arise from terms in Eq.~\eqref{eq:break}. 
}
\label{f:break}
\end{figure}

For the magnetic moments of the $s_\ell = 0$ baryons, the leading light-quark mass dependence
arises at next-to-leading order in the heavy quark expansion. Terms contributing to the 
magnetic moments are due to the mass-splitting of the $B$ and $B^*$ baryons~\cite{Savage:1994zw}, as 
well as from vertices generated from spin-symmetry breaking operators~\cite{Banuls:1999mu}. 
Both of these effects break the $U(2 N_h)$ heavy-quark symmetry and are described in \PQCPT\ 
by the operators
\begin{equation}
\mathfrak{L} 
=  
\lambda \frac{\L^2_{QCD}}{m_Q} 
\left( \ol \cS {}^\mu i \sigma_{\mu \nu} \cS^\nu \right)
- 
\sqrt{2} \lambda_g \frac{\L_{QCD}}{m_Q}
\left[
\left(
\ol \cT A^\mu i \sigma_{\mu \nu} \cS^\nu 
\right)
+
\left(
\ol \cS {}^\mu i \sigma_{\mu \nu} A^\nu \cT
\right)
\right]\label{eq:break}
,\end{equation}
where we have only kept terms relevant for our calculation. The dimensionless parameters $\l$ and $\l_g$ have the same 
numerical values as those in $SU(3)$ \CPT, see Appendix~\ref{cpt}. The terms in this Lagrangian
lead to new vertices that contribute to the magnetic moments of the $s_\ell = 0$ baryons. 
Diagrams involving these heavy quark symmetry breaking vertices are shown in Fig.~\ref{f:break}.

Aside from these loops, we have additionally contributions from the local operators in Eqs.~\eqref{eq:hqmmT} and \eqref{eq:mmT}. Thus
at next-to-leading order in the heavy quark expansion, we find the Pauli form factors of the $s_\ell = 0$ baryons
to be
\begin{eqnarray}
F_2(q^2) &=& 
\frac{M_T}{m_Q} \mu_Q 
+ 
\mu_T \frac{\L_{QCD} M_T}{\L_\chi m_Q} \cQ_T 
+ \l_g
\frac{g_3 \, \L_{QCD} M_T}{2 \pi^2 f^2 \, m_Q}
\sum_\phi \a_\phi^T \int_0^1 dx \, \cF(m_\phi P_\phi, \D, \mu ) 
\notag \\
&& + \l
\frac{g_3^2 \, \L^2_{QCD} M_T}{4 \pi^2 f^2 \, m_Q} \sum_\phi \a_\phi^T \int_0^1 dx  
\Big[ 
\cG(m_\phi P_\phi, \D, \mu) + 2 \log P_\phi + 1
\Big] \label{eq:F2T}
.\end{eqnarray}
The new non-analytic function $\cF(m, \d, \mu)$ entering above
is given by
\begin{equation}
\cF(m, \d, \mu) 
= 
- \d \log \frac{m^2}{\mu^2} 
+
\sqrt{\d^2 - m^2} 
\log 
\frac{\d - \sqrt{\d^2 - m^2 + i \varepsilon}}{\d + \sqrt{\d^2 - m^2 + i \varepsilon}}
\label{eq:Ffn}
.\end{equation}
The coefficients $\a_\phi^T$ are the same as for $F_1(q^2)$ appearing in Table~\ref{t:TPQQCD-A}
and hence the charges can be adjusted in the same manner to simplify the chiral extrapolation
or to decrease the lattice simulation time.

\section{Electromagnetic properties of the $s_{\ell} = 1$ baryons} \label{six}

In this section, we calculate the charge radii, magnetic moments and electric quadrupole moments of the $s_\ell = 1$ baryons 
in \PQCPT. Similar to the above decomposition for the $s_\ell = 0$ baryons, the current 
matrix elements of the spin-$\frac{1}{2}$ $B$ baryons have a decomposition 
\begin{equation} \label{eq:Bdecomp}
\langle \ol B(p') | J^\mu | B(p) \rangle 
= 
\ol u (p') 
\left[ 
v^\mu F_1(q^2)
+
\frac{i \sigma^{\mu \nu} q_\nu}{2 M_B} F_2(q^2)
\right]
u(p)
,\end{equation}
where $q = p' - p$ as the momentum transfer. 
The spin-$\frac{3}{2}$ matrix elements of the electromagnetic current $J^\rho$ for the $B^*$ baryons can be parametrized as
\begin{equation}
\langle \ol B {}_\mu^* (p') | J^\rho | B^*_\nu(p) \rangle 
= 
- \ol u_\mu (p') \cO^{\mu \rho \nu} u_\nu (p)
,\end{equation}
where $u_\mu(p)$ is a Rarita-Schwinger spin vector for an on-shell heavy baryon. This spin vector
satisfies the constraints $v^\mu  u_\mu (p) = \gamma^\mu  u_\mu (p) = 0$.  The tensor $\cO^{\mu \rho \nu}$
can be written in terms of four independent form factors
\begin{equation} \label{eq:B*decomp}
\cO^{\mu \rho \nu} 
=
g^{\mu \nu} 
\left[
v^\rho F^*_1(q^2) + \frac{ i \sigma^{\rho \tau} q_\tau}{2 M_B} F^*_2 (q^2)
\right]
+
\frac{q^\mu q^\nu}{(2 M_B)^2}
\left[
v^\rho G^*_1(q^2) + \frac{i \sigma^{\rho \tau} q_\tau}{2 M_B} G^*_2(q^2) 
\right]
.\end{equation} 
To avoid notational confusion within this section, we have appended a superscript $*$ to denote 
the $B^*$ baryon form factors. 
Extraction of these form factors for the $B^*$ baryons requires 
a non-trivial identity for on-shell Rarita-Schwinger spinors~\cite{Nozawa:1990gt}. 
For our purposes, the identity takes the form
\begin{equation}
\ol u_\a (p') ( q^\a g^{\mu \b} - q^\b g^{\mu \a}) u_\b(p)
= 
\ol u_\a(p') \left(
- \frac{q^2}{2 M_B} g^{\a \b} v^\mu + i g^{\a \b} \sigma^{\mu \nu} q_\nu + \frac{1}{M_B} q^\a q^\b v^\mu  
\right) u_\b(p)
.\end{equation}
The conversion of the above covariant vertex form factors to multipole form factors is detailed 
in~\cite{Nozawa:1990gt}.

The Dirac form factors $F_1(q^2)$ and $F^*_1(q^2)$ for the $B$ and $B^*$ baryons are normalized
at zero momentum transfer to the baryon charge $Q_B$ in units of the electron charge $e$, namely
$F_1(0) = F_1^*(0) = Q_B$. The baryon charge is a sum of the heavy quark charge $\cQ_Q$, and the total charge of 
the light degrees of freedom $\cQ_S$, i.e.~$Q_B = \cQ_Q + \cQ_S$.
Keep in mind that the light-quark
electric charge matrix leaves the valence quarks with arbitrary charges, hence $Q_S$ and $\cQ_S$ need not
have their physical values. The Dirac charge radii for  
have the form $<r^2> = 6 F_1^\prime(0)$, for the $B$ baryons, and $<r^{*2}> = 6 F_1^{*\prime}(0)$, for 
the $B^*$ baryons. 
The magnetic moments of the $B$ and $B^*$ baryons are defined through their Pauli form factors: $\mu = F_2(0)$
and $\mu^* = F^*_2(0)$, respectively. The electric quadrupole moments $\mathbb{Q}$ of the $B^*$ baryons are given by
\begin{equation}
\mathbb{Q} = - \frac{1}{2} G^*_1(0)
.\end{equation}
To the order we work in the heavy quark expansion, the form factor $G^*_2(q^2) = 0$ because the light degrees of freedom have $s_\ell = 1$.
Hence the magnetic octupole moments vanish; see Appendix~\ref{s:HQ} for further discussion. The relations between the electromagnetic 
form factors of the $B$ and $B^*$ baryons are also discussed in Appendix~\ref{s:HQ}.

The electromagnetic observables of the $B$ and $B^*$ baryons receive short and long range contributions in the chiral 
effective theory. The short range contributions stem from both the electromagnetic interaction of the heavy quark 
and the electromagnetic interaction of the light degrees of freedom.
As above, the HQET part arises from the magnetic moment of the heavy quark as well as the electromagnetic Darwin term. 
These contributions are encoded in the effective theory through the heavy quark magnetic moment operator
\begin{equation}
\mathfrak{L} = \frac{e \mu_Q}{4 m_Q} \left( \ol \cS {}^\a  \sigma_{\mu \nu} \cS_\a \right) F^{\mu \nu} 
,\end{equation}
where $\mu_Q = \cQ_Q + \frac{2 \a_s(m_Q)}{3 \pi} + \ldots$, and the heavy quark charge radius operator
\begin{equation}
\mathfrak{L} = \frac{e \cQ_Q}{8 m_Q^2} \left( \ol \cS {}^\a \cS_\a \right) v_\mu \partial_\nu F^{\mu \nu}
.\end{equation}
The electric quadrupole moments of the $B^*$ baryons cannot receive any contribution from the electromagnetic 
interactions of the heavy quark.

In \PQCPT\ the leading contribution from the light degrees of freedom to the magnetic moments of the $B$ and $B^*$ baryons
is contained in the Lagrangian 
\begin{equation}
\mathfrak{L} = \frac{i e \mu_S }{\L_\chi} \left( \ol \cS_\mu \cQ \cS_\nu \right) F^{\mu \nu}
.\end{equation}
To be complete to the order we work, the LEC $\mu_S$ must be treated as a linear function of $\D$, 
i.e.~$\mu_S = \mu_S^{(0)} + \frac{\D}{\L_\chi} \mu_S^{(1)}$. Moreover we absorb the finite contributions
from loop integrals into the parameter $\mu_S^{(1)}$. Determination of this LEC requires the ability to vary $\D$ and
for this reason we treat the $\D$ dependence only implicitly. The magnetic moment coefficient $\mu_S$ is 
the only LEC that must be treated in this fashion to the order we are working.
The charge radii receive local contributions from the dimension-six operator contained in the term
\begin{equation}
\mathfrak{L} = \frac{e c_S }{\L_\chi^2} \left( \ol \cS {}^\a  \cQ  \cS_\a \right) v_\mu \partial_\nu F^{\mu \nu}
,\end{equation}
and finally the electric quadrupole moments arise from 
\begin{equation}
\mathfrak{L} = - \frac{e \mathbb{Q}_S }{\L_\chi^2} \left( \ol \cS {}^{\{\mu}  \cQ \cS^{\nu \}} \right) v_\a \partial_\mu F_{\nu}{}^{ \a}
.\end{equation}
The brackets $\{ \ldots \}$ produce the symmetric traceless part of Lorentz tensors, 
i.e.~$\cO^{\{ \mu \nu \}} = \cO^{\mu \nu} + \cO^{\nu \mu} - \frac{1}{2} g^{\mu \nu} \cO^\a {}_\a$.
The magnetic octupole moment operator enters at $\cO(1/ m_Q \L_\chi^3)$ which is beyond the order we work. 
Although the quark electric charges are arbitrary, 
the electromagnetic LECs for the $s_\ell = 1$ baryons, $\mu_S$, $c_S$ and $\mathbb{Q}_S$, have the same numerical 
values as in $SU(3)$ \CPT, see Appendix~\ref{cpt}.

\begin{figure}
\epsfig{file=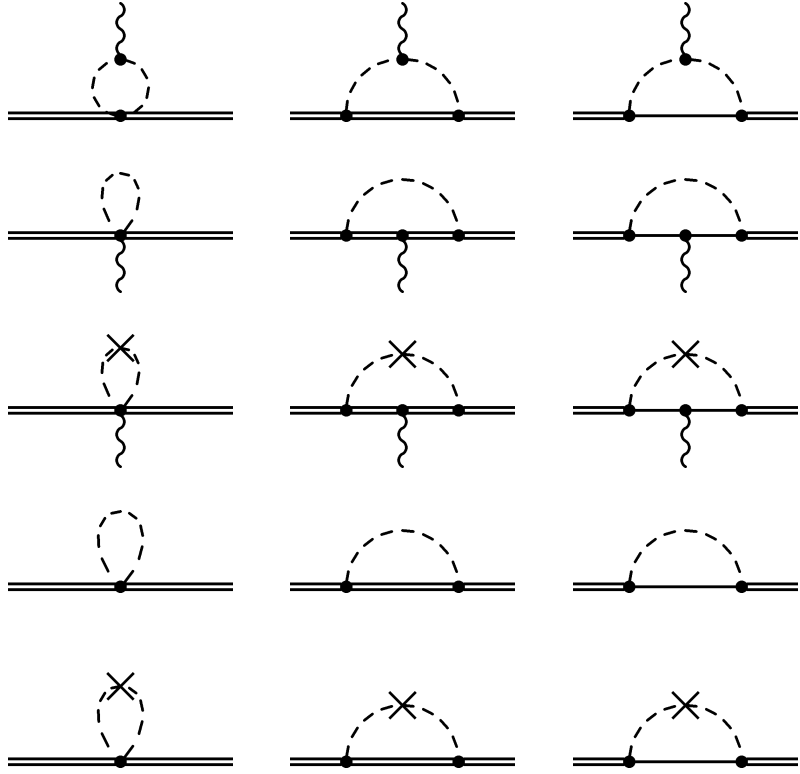}
\label{F:Sloops}
\caption{Loop diagrams contributing to the electromagnetic moments and charge radii of the $s_\ell =1$ baryons. 
Mesons are denoted by a dashed line, flavor neutrals (hairpins) by a crossed dashed line, and the photon by a wiggly line.  
A thick (thin) solid line denotes an $s_\ell = 1$  ($s_\ell =0$) baryon. The diagrams in the first row contribute to the 
electromagnetic moments and charge radii. The remaining diagrams with a photon have no $q^2$-dependence. These, along with the
wavefunction renormalization diagrams in the last two rows, ensure non-renormalization of the electric charge.}
\end{figure}

\begin{table}
\caption{The coefficients $\a^B_\phi$ in $SU(6|3)$ \PQCPT. 
Coefficients are listed for the $s_\ell = 1$ baryon states $B$, and are grouped into contributions from loop mesons
with mass $m_\phi$.}
\begin{tabular}{l | c c c c }
    & $ \qquad ju \qquad \;$ & $ \qquad ru \qquad \;$ 
    & $\qquad js \qquad \;$  & $\qquad rs \qquad \;$ \\
\hline
$\Sigma^{+1}_Q$     
           &  $2 q_u + q_r$ & $q_u - q_r$  
           &  $0$ & $0$ \\
$\Sigma^{0}_Q$     
           &  $q_u + q_d + q_r$ & $\frac{1}{2}(q_u + q_d) - q_r$  
           &  $0$ & $0$ \\
$\Sigma^{-1}_Q$     
           &  $ 2 q_d + q_r$ & $q_d - q_r$  
           &  $0$ & $0$ \\
$\Xi^{\prime + \frac{1}{2}}_Q$     
           &  $q_u + \frac{1}{2} q_r$ & $\frac{1}{2} (q_u - q_r)$  
           &  $q_s + \frac{1}{2} q_r$ & $\frac{1}{2} (q_s - q_r)$ \\
$\Xi^{\prime - \frac{1}{2}}_Q$     
           &  $q_d + \frac{1}{2} q_r$ & $\frac{1}{2} (q_d - q_r) $  
           &  $q_s + \frac{1}{2} q_r$ & $\frac{1}{2} (q_s - q_r) $ \\
$\Omega_Q$     
           &  $0$ & $0$  
           &  $2 q_s + q_r$ & $q_s - q_r$ \\
\end{tabular}
\label{t:SPQQCD-A}
\end{table}

The long distance contributions to the electromagnetic observables of the $B$ and $B^*$ baryons arise 
from the one-loop diagrams depicted in Fig.~\ref{F:Sloops}. These diagrams involve the pseudo-Goldstone mesons 
which enter from the vertices generated by the leading-order interaction Lagrangian Eq.~\eqref{eq:STM}. 
Calculation of these diagrams along with the tree-level contributions for the $B$ baryons yields
\begin{eqnarray}
F_1(q^2) 
&=&
Q_B + 
\left(
\cQ_Q \frac{1}{8 m_Q^2} + \cQ_S \frac{c_S}{\L_\chi^2} + \cQ_S \frac{\mathbb{Q}_S}{6 \L_\chi^2}
\right) q^2
\notag \\
&& +
\frac{1}{8 \pi^2 f^2} 
\sum_\phi \a_\phi^B 
\int_0^1 dx
\left[ 
-\frac{q^2}{6} \log \frac{m_\phi^2}{\mu^2} 
+
2 m_\phi^2 P_\phi^2 \log P_\phi
\right]
\notag \\
&& + 
\frac{g_2^2}{(4 \pi f)^2}
\sum_\phi \a_\phi^B
\int_0^1 dx
\Bigg\{
2 \left[
m_\phi^2 - \frac{5}{3} x (1-x) q^2
\right]
\log P_\phi
- 
\frac{5}{18} q^2 
\left[ 
\log \frac{m_\phi^2}{\mu^2} + 1
\right]
\Bigg\}
\notag \\
&& + 
\frac{g_3^2}{(4 \pi f)^2} \sum_\phi \a_\phi^B
\int_0^1 dx 
\Bigg[
\cJ(m_\phi P_\phi, -\D, \mu) - \cJ(m_\phi, -\D, \mu)
+ \frac{1}{9} q^2
\notag \\
&& \phantom{morespacesmorespaces}
- \frac{2}{3} x (1-x) q^2 \,
\cG (m_\phi P_\phi, -\D, \mu)  
\Bigg] \label{eq:F1B}
\end{eqnarray}
for the Dirac form factor and
\begin{eqnarray}
F_2(q^2) 
&=&
\frac{M_B}{m_Q} \mu_Q + \mu_S \frac{4 M_B}{3 \L_\chi} \cQ_S 
+ 
\frac{g_2^2 M_B}{24 \pi f^2} 
\sum_\phi \a_\phi^B
\int_0^1 dx \, m_\phi P_\phi
\notag \\
&& + 
\frac{g_3^2 M_B}{12 \pi^2 f^2} 
\sum_\phi \a_\phi^B 
\int_0^1 dx \,
\cF(m_\phi P_\phi, -\D, \mu)
\label{eq:F2B}
\end{eqnarray}
for the Pauli form factor. The function $\cF(m, \d, \mu)$ appearing above is given in Eq.~\eqref{eq:Ffn}. 
In the above expressions, the sums run over loop mesons $\phi$ with mass $m_\phi$. The coefficients $\a_\phi^B$
are listed in Table~\ref{t:SPQQCD-A} and depend on the particular baryon state $B$. 
Notice the charges $q_j$ and $q_l$ do not explicitly appear in these coefficients.  
In the isospin limit, they always enter in the combination $q_j + q_l$ which is the same as $-q_r$ by supertracelessness of $\cQ$.

Carrying out the calculation for the $B^*$ baryons, we obtain
\begin{eqnarray}
F_1^*(q^2) 
&=&
Q_B + 
\left(
\cQ_Q \frac{1}{8 m_Q^2} + \cQ_S \frac{c_S}{\L_\chi^2} - \cQ_S \frac{\mu_S}{2 \L_\chi M_B}- \cQ_S \frac{\mathbb{Q}_S}{2 \L_\chi^2}
\right) q^2
\notag \\
&& +
\frac{1}{8 \pi^2 f^2} 
\sum_\phi \a_\phi^B 
\int_0^1 dx
\left[ 
-\frac{q^2}{6} \log \frac{m_\phi^2}{\mu^2} 
+
2 m_\phi^2 P_\phi^2 \log P_\phi
\right]
\notag \\
&& + 
\frac{g_2^2}{(4 \pi f)^2}
\sum_\phi \a_\phi^B
\int_0^1 dx
\Bigg\{
2 \left[ m_\phi^2 - 2 x(1-x) q^2 \right]
\log P_\phi
- \frac{1}{3} q^2 \left[ \log \frac{m_\phi^2}{\mu^2} + 1 \right]
\Bigg\}
\notag \\
&& + 
\frac{g_3^2}{(4 \pi f)^2} \sum_\phi \a_\phi^B
\int_0^1 dx 
\Big[
\cJ(m_\phi P_\phi, -\D, \mu) - \cJ(m_\phi, -\D, \mu)
\Big]
\label{eq:F1B*}
\end{eqnarray}
for the Dirac form factor, 
\begin{eqnarray}
F_2^*(q^2) 
&=&
\frac{M_B}{m_Q} \mu_Q + \mu_S \frac{2 M_B}{\L_\chi} \cQ_S 
+ 
\frac{g_2^2 M_B}{16 \pi f^2}
\sum_\phi \a_\phi^B
\int_0^1 dx 
m_\phi P_\phi
\notag \\
&& + 
\frac{g_3^2 M_B}{8 \pi^2 f^2} 
\sum_\phi \a_\phi^B 
\int_0^1 dx \,
\cF(m_\phi P_\phi, -\D, \mu)
\label{eq:F2B*}
\end{eqnarray}
for the Pauli form factor, and finally
\begin{eqnarray}
G_1^*(q^2) 
&=& 
4 \cQ_S \frac{M_B}{\L_\chi} 
\left(
\mu_S + 2 \mathbb{Q}_S \frac{M_B}{\L_\chi}
\right)
\notag \\
&& +
\frac{g_2^2 M_B^2}{4 \pi^2 f^2}
\sum_\phi \a_\phi^B
\int_0^1 dx 
\left\{
2 x (1- x) \log P_\phi
+
\frac{1}{6}
\left[
\log \frac{m_\phi^2}{\mu^2}
+ 1
\right]
\right\}
\notag \\
&& -
\frac{g_3^2 M_B^2}{2 \pi^2 f^2}
\sum_\phi \a_\phi^B
\int_0^1 dx 
\Big[ 
x (1-x) \, \cG(m_\phi P_\phi, -\D, \mu)
+ \frac{1}{6}
\Big]
\label{eq:G1B*}
.\end{eqnarray}
The coefficients $\a_\phi^B$ appearing in the expressions for the $B^*$ baryon form factors
are listed in Table~\ref{t:SPQQCD-A} and are identical for baryon states $B$ and $B^*$. 
Taking the $m_Q \to \infty$ limit, the relations between the $B$ and $B^*$ electromagnetic form 
factors (derived in Appendix~\ref{s:HQ}) are satisfied by our one-loop \PQCPT\ results. 

As with the $s_\ell = 0$ baryons, Table~\ref{t:SPQQCD-A}
shows that the chiral extrapolations can be simplified with efficacious choices for the charges of valence and sea quarks.
For the $\Sigma_Q$ and $\Omega_Q$ baryons, we can eliminate all but one loop meson; while for the $\Xi^\prime_Q$ baryons,
we can reduce the contributions from four loop mesons down to two. 
In $SU(6|3)$ one can again decrease the computation time by choosing the sea quarks to have vanishing electric charges 
as this preserves the supertracelessness of the charge matrix. Thus all closed quark loops with a current insertion vanish. 
Furthermore for the $\Sigma^0_Q$, and $\Xi^\prime_Q$ baryons, either of the contributing valence charges can be taken to zero, 
and the LECS can still be extracted from the lattice data. For the $\Sigma^{+1}_Q$, $\Sigma^{-1}_Q$, 
and $\Omega_Q$ baryons, this choice is not possible since these states have two valence quarks of the same flavor. 
Nonetheless, LECs extracted from calculations involving one baryon can be used to test \PQCPT\ predictions
for the remaining baryons.

\section{Summary} \label{summy}

Above we have calculated the electromagnetic properties of baryons containing a heavy quark. 
For baryons with light degrees of freedom $s_\ell = 0$, we obtained the charge radii
at NLO in the chiral expansion. The magnetic moments of these baryons vanish at leading order
in the heavy-quark expansion and thus to obtain the leading light-quark mass dependence
we worked to NLO in the heavy quark expansion. For the $s_\ell = 1$ baryons, we obtained
the charge, magnetic moment and electric quadrupole moment form factors at NLO in the 
chiral expansion. Expressions for these quantities are derived in \PQCPT\ for three 
light flavors in the main text, and two light flavors in Appendix~\ref{pqsutwo}. 
Additionally the corresponding \CPT\ results for three and two light flavors 
are presented in Appendix~\ref{cpt}. Knowledge of the quark mass dependence of these
observables is essential to extrapolate lattice QCD data to the physical light quark masses. 
Additionally the effective field theory predicts the momentum transfer dependence 
of the electromagnetic form factors near zero recoil. This information should be utilized to perform the 
zero-momentum extrapolation of lattice data in order to extract static electromagnetic properties.
Furthermore there are relations between the various sextet baryon electromagnetic
form factors that follow from the decoupling of the heavy and light spins; 
these are found in Appendix~\ref{s:HQ}.

There are a number of interesting points that arise when considering the electromagnetic 
properties of baryons containing a heavy quark as compared to baryons containing only light quarks. 
Firstly the choice of the heavy quark lattice action allows for more freedom in separating contributions
to baryon properties. For the octet and decuplet baryons formed from three light quarks, in order 
to have a consistent power counting the average inverse baryon mass $1/M$ becomes an expansion 
parameter~\cite{Jenkins:1991jv,Jenkins:1991es}. On the lattice, however, there is no direct control
over this expansion, whereas the choice of the HQET action leads to control over the $1/m_Q$
expansion. This in turn can provide information about the chiral regime. Consider
the $s_\ell = 0$ baryons. The form factors $F_1(q^2)$ and $F_2(q^2)$ enter at different orders in 
the heavy-quark expansion. Thus if the light-quark mass dependence from \CPT\ can be trusted, then
increasing the heavy quark mass should diminish the variation of $F_2(q^2)$ and isolate the leading 
loop contributions to $F_1(q^2)$. If this behavior is not observed, one is certainly not in the chiral regime.

A further contrast to the baryons with only light quarks, is the case of the $\L_Q$ and $\Sigma_Q^0$
baryons. In $SU(2)$ flavor, the loop contributions to the electromagnetic properties of these baryons 
vanish due to cancellations between $\pi^+$ and $\pi^-$ loops. Such complete cancellations do 
not occur for the nucleons and deltas.\footnote{%
There is the case of the spin-$\frac{3}{2}$ $\Sigma^{*,0}$ baryon in $SU(3)$ for which the pion and kaon 
loops individually sum to zero. The local electromagnetic operators, however, are all proportional 
to the baryon charge, thus to one-loop order all electromagnetic observables of the $\Sigma^{*,0}$ vanish. 
}  
From the point of view of chiral physics, this is perhaps uninteresting as there is no quark mass dependence.
This situation is true at NLO in the heavy quark expansion as well, because the flavor structure of 
the heavy quark symmetry breaking operators is identical to the LO operators. If there is observed quark mass 
variation of the electromagnetic properties of the $\L_Q$ and $\Sigma_Q^0$, then one is not in the chiral regime. 
Even though there is no quark-mass dependence in QCD, the couplings $g_2$ and $g_3$ can still be extracted 
from considering the $\L_Q$ and $\Sigma_Q^0$ in \PQCPT, because the arbitrary quark electric charges can be chosen
so that loop diagrams do contribute.

A feature of three-flavor electromagnetism, is the tracelessness of the charge matrix. This 
carries over into the partially quenched theory as supertracelessness, and enables the LECs to be extracted
from lattice data in which the sea quarks have vanishing charges. This reduces the computation time, as closed
quark loops with current insertion vanish. We have accordingly used a charge matrix which clearly
separates valence and sea contributions; see Appendix~\ref{s:charge}. Similar electroweak operators 
should be utilized for other observables.

Lattice QCD calculations will enable an understanding of the electromagnetic properties 
of hadrons in terms of quark and gluon degrees of freedom. These observables 
paint an intuitive picture of the low-energy electromagnetic structure of hadrons. 
Foreseeable lattice calculations of these properties, however, will require
extrapolation in the light quark masses. The expressions derived in this paper for heavy hadrons in
\CPT\ and \PQCPT\ allow such extrapolations, as well as the zero-momentum extrapolation.
The \PQCPT\ expressions are not complicated by new non-analytic functions arising
from the flavor-neutral propagator. Moreover a study of the electromagnetic properties 
of baryons containing a heavy quark allows the exploration of our ability to extrapolate 
lattice QCD on two fronts: the light-quark regime and the heavy-quark regime. 
Control over the latter scale is not a feature for nucleons and deltas treated as heavy baryons. 
Lastly the exploitation of arbitrary electroweak quark charges allows for calculational simplifications
to extract physical low-energy constants.

\acknowledgments
We thank Tom Mehen for helpful discussions 
and Andr\'e Walker-Loud for correspondence. 
This work was supported in part by the U.S. 
Department of Energy under Grant No.~DE-FG02-96ER40945.

\appendix

\section{Electromagnetic properties in $SU(4|2)$} \label{pqsutwo}

Heavy hadron chiral perturbation theory in the 
baryon sector for $SU(4|2)$ parallels that of $SU(6|3)$ 
in Sec.~\ref{pqhhcpt}. The partially quenched heavy hadron Lagrangian 
has been written down in~\cite{Arndt:2003vx}. In this appendix, we
briefly review the setup of $SU(4|2)$ PQQCD and \PQCPT. We then 
present the calculation of heavy hadron electromagnetic properties 
for this graded flavor group.

\subsection{\PQCPT\ for $SU(4|2)$}

In $SU(4|2)$ PQQCD, the light-quark sector Lagrangian appears as
\begin{equation}
\mathfrak{L} = \sum_{j,k=1}^6 \ol{q}_j \left(
  i\Dslash - m_q \right)_{jk} q_k
\label{eq:pqqcdlag2}
,\end{equation}
and differs from the usual QCD Lagrangian by the inclusion of four extra quarks; 
two bosonic ghost quarks, ($\tilde u, \tilde d$), and two fermionic sea quarks, ($j, l$), 
in addition to the light physical quarks ($u, d$).  The six quark fields transform in the 
fundamental representation of $SU(4|2)$.  
They appear in the six-component vector
\begin{equation}
q = (u, d, j, l,\tilde{u}, \tilde{d})^{\text{T}}
.\end{equation}
The quark fields obey the graded equal-time commutation relations in Eq.~\eqref{eq:qetcr}
but with the grading factor $\eta_k$ now defined by
\begin{equation}
   \eta_k
   = \left\{ 
       \begin{array}{cl}
         1 & \text{for } k=1,2,3,4 \\
         0 & \text{for } k=5,6
       \end{array} 
     \right.
.\end{equation}
The $SU(4|2)$ mass matrix with non-degenerate quarks is given by
\begin{equation}
m_q = \diag(m_u, m_d, m_j, m_l, m_u, m_d)
,\end{equation}
where the ghost quarks remain degenerate with their valence partners.
Defining ghost and sea quark charges is constrained only be the restriction that QCD be recovered
in the limit of appropriately degenerate quark masses. The general 
form of the charge matrix we choose is
\begin{equation} \label{eq:chargeSU2}
\cQ = \diag 
\left(
q_u, q_d, q_j, q_l , q_u , q_d 
\right)
,\end{equation}
which is not supertraceless. In the limit $m_j \rightarrow m_u$ and $m_l \rightarrow m_d$, 
QCD with two light flavors is recovered only for the values $q_u = q_j = \frac{2}{3}$, and
$q_d = q_l = - \frac{1}{3}$. The results for and problems with the commonly used partially-quenched charge matrix 
are presented in Appendix~\ref{s:charge}.

For massless light quarks, 
the Lagrangian in Eq. (\ref{eq:pqqcdlag2}) has a graded symmetry 
$SU(4|2)_L \otimes SU(4|2)_R \otimes U(1)_V$, which is
assumed to be spontaneously broken to 
$SU(4|2)_V \otimes U(1)_V$.  
The Lagrangian describing the pseudo-Goldstone mesons of this theory
is identical in form to Eq.~\eqref{eq:pqbosons}. The meson fields appearing in Eq.~\eqref{eq:mesonmatrix} are replaced by 
an $SU(4|2)$ matrix $\Phi$. 
The upper $2 \times 2$ block of the matrix $M$ contains the familiar
pions, see e.g.~\cite{Beane:2002vq}.
The flavor singlet field that appears in $SU(4|2)$  is defined to be 
$\Phi_0 = {\rm str}( \Phi ) / {\sqrt 2}$.  As before, 
the mass of the singlet field $m_0$ can be taken to be
on the order of the chiral symmetry breaking scale.  
In this limit, the $\eta$ two-point correlation functions deviate from their familiar form in \CPT,
and their explicit forms are not needed for expressing our results.

To include baryons with one heavy quark into the theory, we use the same interpolating
fields, Eqs.~\eqref{eq:Tinterp} and \eqref{eq:Sinterp}.
The $s_\ell = 0$ baryons are then described by the field $\cT_{ij}$
which forms a $\mathbf{17}$-dimensional representation of $SU(4|2)$. 
The baryon tensor of QCD $T_{ij}$ is contained as $\cT_{ij} = T_{ij}$, 
when the indices are restricted to the range $1-2$. 
In our conventions, we have 
\begin{equation}
T_{ij} 
= 
\frac{1}{\sqrt{2}}
\begin{pmatrix}
0              &     \L_Q     \\ 
- \L_Q         &        0     
\end{pmatrix}_{ij}
\label{eq:TSU2}.\end{equation}
The $s_\ell = 1$ baryons are described by $\cS^\mu_{ij}$
which makes up a $\mathbf{19}$-dimensional representation of $SU(4|2)$. 
The baryon tensor of QCD $S^\mu_{ij}$ is embedded as 
$\cS^\mu_{ij} = S^\mu_{ij}$, when the indices are restricted to the range $1-2$.
Here the QCD flavor tensor $S^\mu_{ij}$ is given in terms of $B_{ij}$ and $B^*_{ij}$ as in Eq.~\eqref{eq:Sdecomp}, but with
\begin{equation}
B_{ij} 
= 
\begin{pmatrix}
\Sigma_Q^{+1}                  &  \frac{1}{\sqrt{2}} \Sigma_Q^0  \\ 
\frac{1}{\sqrt{2}} \Sigma_Q^0  &  \Sigma_Q^{-1}                   
\end{pmatrix}_{ij}
\label{eq:BSU2},\end{equation}
and similarly for $B^{*\mu}_{ij}$. The superscript on these states labels the $3$-projection of isospin. 
The remaining states relevant to our calculation have been classified in~\cite{Arndt:2003vx}.

The free Lagrangian for the $\cT$ and $\cS^\mu$ fields is the same as Eq.~\eqref{eq:STfree}
and the Lagrangian that describes the interactions of these fields with 
the pseudo-Goldstone modes is given by Eq.~\eqref{eq:STM}. 
The LECs appearing in the Lagrangian, $\l_1$, $\l_2$, $g_2$, and $g_3$, all have 
the same numerical values as those used in $SU(2)$ heavy hadron \CPT, 
see Appendix \ref{cpt}. These are of course numerically different than those in $SU(3)$. 
The parameters $\l_3^{(PQ)}$ and $\l_4^{(PQ)}$ are different and can be related to the unquenched coefficient
$\l_4$ by matching, namely $\l_4 = \frac{1}{2} \l_3^{(PQ)} + \l_4^{(PQ)}$.

\subsection{Baryon electromagnetic properties}
Calculation of the baryon electromagnetic properties in $SU(4|2)$ is similar
to that in $SU(6|3)$ because the Lagrangian has the same structure. There is, however,  one crucial 
difference because the charge matrix is not supertraceless and hence there are new operators
contributing at tree level to each of the electromagnetic properties.

For the Dirac form factor of the $s_\ell = 0$ baryons, there is an additional
operator that contributes at tree level, namely
\begin{equation}
\mathfrak{L} = - 
\frac{e \ol c_T}{\L_\chi^2} 
\left( \ol \cT \cT \right)
v_\mu \partial_\nu F^{\mu \nu} \, \str \cQ
.\end{equation}
The contribution of this operator to the Dirac form factor of the $\L_Q$ baryon is
\begin{equation}
\d F_1(q^2) = \frac{q_{jl} \, \ol c_T}{\L_\chi^2} q^2 
,\end{equation}
where $q_{jl} = q_j + q_l$ is the supertrace of the charge matrix.  
The remainder of the form factor is given in Eq.~\eqref{eq:F1T}, however the coefficients $\a^T_\phi$
are now listed for the $\L_Q$ baryon for $SU(4|2)$ in Table~\ref{t:TPQQCD-2}. 
For the magnetic moment of the $\L_Q$, we have the additional operator
\begin{equation}
\mathfrak{L} = - \frac{e \ol \mu_T \L_{QCD}}{4 \L_\chi m_Q} 
\left( \ol \cT \sigma_{\mu \nu} \cT \right) 
\, F^{\mu \nu} \, \str \cQ 
,\end{equation}
which leads to a contribution to the Pauli form factor
\begin{equation}
\d F_2(q^2) = q_{jl} \, \ol \mu_T \frac{\L_{QCD} M_T}{\L_\chi m_Q}
,\end{equation}
and the remainder of the Pauli form factor is given in Eq.~\eqref{eq:F2T} with the coefficients $\a_\phi^T$ for $SU(4|2)$.

\begin{table}
\caption{The coefficients $\a^T_\phi$ for the $\Lambda_Q$ in $SU(4|2)$ \PQCPT. 
Coefficients  are grouped into contributions from loop mesons with mass $m_\phi$.} 
\begin{tabular}{l | c c c c }
    & $ \qquad ju \qquad \;$ & $ \qquad lu \qquad \;$ 
    & $\qquad jd \qquad \;$  & $\qquad ld \qquad \;$ \\
\hline
$\Lambda_Q$     
           &  $ \frac{1}{2} (q_u - q_{j})$ & $\frac{1}{2} (q_u - q_l)$  
           &  $ \frac{1}{2} (q_d - q_j)$ & $\frac{1}{2} (q_d - q_l)$ \\
\end{tabular}
\label{t:TPQQCD-2}
\end{table}

From Table~\ref{t:TPQQCD-2}, we see that the charges of the valence and sea quarks can be chosen so that there 
are no loop contributions at this order. This dramatically simplifies the partially quenched chiral extrapolation
and can be anticipated from the $SU(2)$ result, see Appendix~\ref{cpt}. 
One can choose $\cQ$ to be supertraceless to sort out $c_T$ from $\ol c_T$, and $\mu_T$ from $\ol \mu_T$. 
The specific choice $q_j = q_l = 0$ is possible, but the result is not sensitive to all the LECs. Thus in $SU(4|2)$
quark disconnected contributions with current insertion must be dealt with to ascertain all LECs and hence make physical 
predictions.

Considering next the $s_\ell = 1$ baryon electromagnetic properties, we must add similar types of operators involving
the supertrace of the charge matrix. These operators are
\begin{equation}
\mathfrak{L} = 
\frac{i e \ol \mu_S}{\L_\chi} 
\left( \ol \cS_\mu  \cS_\nu \right) \, F^{\mu \nu} \, \str \cQ
+ 
\frac{e \ol c_S}{\L_\chi^2}
\left( \ol \cS {}^\a \cS_\a \right) \, v_\mu \partial_\nu F^{\mu \nu} \, \str \cQ
- 
\frac{e \overline{\mathbb{Q}} {}_S }{ \L_\chi^2 } 
\left( \ol \cS {}^{\{ \mu} \cS^{\nu \}} \right) \, v_\a \partial_\mu F_{\nu}{}^{ \a} \, \str \cQ
.\end{equation} 
For the spin-$\frac{1}{2}$ $B$ baryons, the contributions to the form factors from the above 
terms are
\begin{eqnarray}
\d F_1(q^2) &=&  
\left(
\frac{\ol c_S}{\L_\chi^2}
+ 
\frac{\overline{\mathbb{Q}} {}_S}{6 \L_\chi^2}
\right) q_{jl} \, q^2
\notag \\
\d F_2(q^2) &=&  \frac{4 M_B}{3 \L_\chi} q_{jl} \, \ol \mu_S
.\end{eqnarray}
The remaining parts of the Dirac and Pauli form factors are the same as in Eqs.~\eqref{eq:F1B} and
\eqref{eq:F2B}, respectively. The coefficients $\a_\phi^B$ for $SU(4|2)$ are listed in Table~\ref{t:SPQQCD-2} 
for the $\Sigma_Q$ baryons. 
Lastly, for the spin-$\frac{3}{2}$ $B^*$ baryons, the contributions to the electromagnetic 
form factors from the above operators are
\begin{eqnarray}
\d F^*_1(q^2) &=& 
\left(
\frac{\ol c_S}{\L_\chi^2} - \frac{\ol \mu_S }{2 \L_\chi M_B} - \frac{\overline{\mathbb{Q}} {}_S}{2 \L_\chi^2}
\right) q_{jl} \, q^2
\notag \\
\d F^*_2(q^2) &=&
\frac{2 M_B}{\L_\chi} q_{jl} \, \ol \mu_S
\notag \\
\d G^*_1(q^2) &=&
\frac{4 M_B}{\L_\chi} \left( \ol \mu_S +  2 \overline{\mathbb{Q}} {}_S \frac{M_B}{\L_\chi} \right) q_{jl}
,\end{eqnarray}
with the remaining parts of these three form factors given in Eqs.~\eqref{eq:F1B*}, \eqref{eq:F2B*} and \eqref{eq:G1B*}
with the coefficients $\a_\phi^B$ listed in Table~\ref{t:SPQQCD-2}.

\begin{table}
\caption{The coefficients $\a^B_\phi$ in $SU(4|2)$ \PQCPT. 
Coefficients are listed for the $s_\ell = 1$ baryon states $B$, and are grouped into contributions from loop mesons
with mass $m_\phi$.}
\begin{tabular}{l | c c c c }
    & $ \qquad ju \qquad \;$ & $ \qquad lu \qquad \;$ 
    & $\qquad jd \qquad \;$  & $\qquad ld \qquad \;$ \\
\hline
$\Sigma^{+1}_Q$     
           &  $q_u - q_{j}$ & $q_u - q_l$  
           &  $0$ & $0$ \\
$\Sigma^{0}_Q$     
           &  $\frac{1}{2} (q_u - q_{j})$ & $\frac{1}{2} (q_u - q_l)$  
           &  $\frac{1}{2} (q_d - q_j)$ & $\frac{1}{2} (q_d - q_l)$ \\
$\Sigma^{-1}_Q$     
	   &  $0$ & $0$	           
           &  $q_d - q_{j}$ & $q_d - q_l$  \\           
\end{tabular}
\label{t:SPQQCD-2}
\end{table}

From Table~\ref{t:SPQQCD-2}, we see that for each $\Sigma_Q$ baryon the charges of the valence and sea quarks can be chosen 
so that there are no loop contributions at this order. Consequently there is a simple partially-quenched chiral extrapolation.
Compared with the $SU(2)$ result, see Appendix~\ref{cpt}, only the $\Sigma_Q^0$ has this simplification. 
One can choose $\cQ$ to be supertraceless to sort out $c_S$ from $\ol c_S$, etc. 
The specific choice $q_j = q_l = 0$ is possible, but again does not lead to results that depend on all LECs. 
Quark disconnected contributions with current insertion must be calculated in $SU(4|2)$ 
to ascertain the relevant LECs and make physical predictions.

\section{Electromagnetic properties in $SU(3)$ and $SU(2)$} \label{cpt}

For completeness, we include calculations of the baryon electromagnetic properties in \CPT. 
The magnetic moments were calculated to one-loop order in~\cite{Savage:1994zw,Banuls:1999mu}; 
the remaining properties have not been calculated before. 
To compare with the main text, we consider the case of $SU(3)$ flavor in the isospin limit
and to compare with the results of Appendix~\ref{pqsutwo}, we consider non-degenerate $SU(2)$. 
For these calculations, we retain the tensors $T_{ij}$ and $S^\mu_{ij}$ in \CPT. For the 
case of $SU(3)$ flavor the tensors are given in Eqs.~\eqref{eq:TSU3} and \eqref{eq:BSU3}, respectively; while
for $SU(2)$ flavor, they are given by Eqs.~\eqref{eq:TSU2} and \eqref{eq:BSU2}, respectively.

The free Lagrangian for the $T_{ij}$ and $S^\mu_{ij}$ fields in $SU(3)$ and $SU(2)$ both have the form
\begin{eqnarray}
\mathfrak{L} 
&=& -
i \left( \ol S {}^\mu v \cdot D S_\mu \right)
+ 
\D \left( \ol S {}^\mu S_\mu \right)
+ 
\l_1 \left( \ol S {}^\mu  \cM S_\mu \right)
+ 
\l_2 \left( \ol S {}^\mu S_\mu \right) \tr \cM
\notag \\
&& + 
i \left( \ol T v \cdot D T \right) 
+ 
\l_3 \left( \ol T \cM T \right)
+ 
\l_4 \left( \ol T T \right) \tr \cM 
\label{eq:STfreeCPT}
,\end{eqnarray}
with the exception that $\l_3 = 0$ in the two-flavor theory to avoid redundancy~\cite{Tiburzi:2004kd}. 
Similarly the interaction Lagrangian for $T_{ij}$ and $S^\mu_{ij}$ has the same form for both flavor groups, 
namely
\begin{equation}
\mathfrak{L} 
= 
i g_2 
\left( 
\ol S {}^\mu v^\nu A^\rho S^\sigma 
\right)
\varepsilon_{\mu \nu \rho \sigma}
+
\sqrt{2} \, g_3 
\left[ 
\left( 
\ol T A^\mu S_\mu 
\right) 
+ 
\left(
\ol S {}^\mu A_\mu T
\right)
\right]
\label{eq:STMCPT}
\end{equation}
The factor $\sqrt{2}$ is chosen so that $g_3$ agrees with the literature, e.g.~\cite{Manohar:2000dt}.
The numerical values of the $\l_j$, $g_2$, and $g_3$ are of course different for the $SU(3)$ and $SU(2)$ theories.

\subsection{$SU(3)$}

In contrast to the partially quenched theories considered above, the light-quark 
mass matrix in the isospin limit of $SU(3)$ flavor is given by
\begin{equation}
m_q = \diag ( m_u, m_u, m_s ),	  
\end{equation}
and the light-quark charge matrix is
\begin{equation}
\cQ = \diag \left( \frac{2}{3} , - \frac{1}{3} , - \frac{1}{3} \right)
.\end{equation}
The local operators contributing to the electromagnetic properties of the $s_\ell = 0$ baryons
are contained in the Lagrangian
\begin{equation}
\mathfrak{L} = 
- \frac{e \mu_Q}{4 m_Q} \left( \ol T \sigma_{\mu \nu} T \right) F^{\mu \nu} 
-  \frac{e \cQ_Q}{8 m_Q^2} \left( \ol T T \right)  v_\mu \partial_\nu F^{\mu \nu}
\label{eq:local1},\end{equation}
for the heavy quark contributions and
\begin{equation}
\mathfrak{L} = 
- \frac{e \mu_T \L_{QCD}}{4 \L_\chi m_Q} \left( \ol T \cQ \sigma_{\mu \nu} T \right) F^{\mu \nu}
- \frac{e c_T }{\L_\chi^2} \left( \ol T \cQ T \right) v_\mu \partial_\nu F^{\mu \nu}
\label{eq:local2},\end{equation}
for the brown muck. The LECs $\mu_T$ and $c_T$ in $SU(6|3)$ have the same numerical values as
in $SU(3)$ \CPT.
The heavy quark symmetry breaking operators have the form
\begin{equation}
\mathfrak{L} 
=  
\lambda \frac{\L^2_{QCD}}{m_Q} 
\left( \ol S {}^\mu i \sigma_{\mu \nu} S^\nu \right)
- 
\sqrt{2} \lambda_g \frac{\L_{QCD}}{m_Q}
\left[
\left(
\ol T A^\mu i \sigma_{\mu \nu} S^\nu 
\right)
+
\left(
\ol S {}^\mu i \sigma_{\mu \nu} A^\nu T
\right)
\right]
\label{eq:local3}\end{equation}
for the $SU(3)$ flavor group. 
We trivially find the LECs above, $\l$ and $\l_g$,  have the same values as in $SU(6|3)$. 
The Dirac and Pauli form factors can now be determined in $SU(3)$ and they have a form 
identical to Eqs.~\eqref{eq:F1T} and \eqref{eq:F2T}. The coefficients $\a_\phi^T$
are listed for $SU(3)$ in Table~\ref{t:TQCD-A}.

\begin{table}
\caption{The coefficients $\a^T_\phi$ in $SU(3)$ \CPT. Coefficients are
listed for the $s_\ell = 0$ baryon states $B$, and are grouped into contributions from loop mesons
with mass $m_\phi$. }
\begin{tabular}{l | r r }
    & $\qquad \pi \quad $ & $\qquad  K \quad $  \\
\hline
$\L_Q$                 &  $0$ & $\frac{1}{2}$   \\
$\Xi_Q^{+\frac{1}{2}}$  &  $\frac{1}{2}$ & $0$   \\
$\Xi_Q^{-\frac{1}{2}}$  &  $-\frac{1}{2}$ & $-\frac{1}{2}$   \\
\end{tabular}
\label{t:TQCD-A}
\end{table}

The local operators contributing to the electromagnetic properties of the $s_\ell = 1$ baryons
are contained in the Lagrangian
\begin{equation}
\mathfrak{L} = 
\frac{e \mu_Q}{4 m_Q} \left( \ol S {}^\a  \sigma_{\mu \nu} S_\a  \right) F^{\mu \nu} 
+ \frac{e \cQ_Q}{8 m_Q^2} \left( \ol S {}^\a  S_\a \right) v_\mu \partial_\nu F^{\mu \nu}
\label{eq:local4},\end{equation}
for the heavy quark contributions and
\begin{equation}
\mathfrak{L} = 
\frac{i e \mu_S}{\L_\chi} \left( \ol S_\mu \cQ S_\nu \right) F^{\mu \nu}
+
\frac{e c_S }{\L_\chi^2} \left( \ol S {}^\a \cQ S_\a \right) v_\mu \partial_\nu F^{\mu \nu}
- 
\frac{e \mathbb{Q}_S}{\L_\chi^2} \left( \ol S {}^{\{\mu} \cQ S^{\nu\}} \right)
v_\a \partial_\mu F_{\nu}{}^{ \a}
\label{eq:local5},\end{equation}
for the brown muck. The LECs $\mu_S$, $c_S$, and $\mathbb{Q}_S$ in $SU(6|3)$ have the same numerical values as
those in $SU(3)$ \CPT. The three form factors $F_1(q^2)$, and $F_2(q^2)$ can now be determined in $SU(3)$ for the $B$
baryons and they have a form identical to Eqs.~\eqref{eq:F1B} and \eqref{eq:F2B}. The form factors $F_1^*(q^2)$, 
$F_2^*(q^2)$ and $G_1^*(q^2)$ for the $B^*$ baryons have the same form as Eqs.~\eqref{eq:F1B*}, \eqref{eq:F2B*} and
\eqref{eq:G1B*}. For each of the form factors for the $B$ and $B^*$ baryons, the coefficients $\a_\phi^B$ are listed 
for $SU(3)$ in Table~\ref{t:SQCD-A}.

\begin{table}
\caption{The coefficients $\a^B_\phi$ in $SU(3)$ \CPT. Coefficients are
listed for the $s_\ell = 1$ baryon states $B$, and are grouped into contributions from loop mesons
with mass $m_\phi$. }
\begin{tabular}{l | r r }
    & $\qquad \pi \quad $ & $\qquad  K \quad $  \\
\hline
$\Sigma^{+1}_Q$               & $1$ & $1$ \\
$\Sigma^{0}_Q$                & $0$ & $\frac{1}{2}$ \\
$\Sigma^{-1}_Q$               & $-1$ & $0$ \\
$\Xi^{\prime +\frac{1}{2}}_Q$ & $\frac{1}{2}$ & $0$ \\
$\Xi^{\prime -\frac{1}{2}}_Q$ & $-\frac{1}{2}$ & $-\frac{1}{2}$ \\
$\Omega_Q$ 	   	      & $0$ & $-1$ \\
\end{tabular}
\label{t:SQCD-A}
\end{table}

\subsection{$SU(2)$}
The light-quark mass matrix in $SU(2)$ flavor is given by
\begin{equation}
m_q = \diag ( m_u, m_d ),	  
\end{equation}
and the light-quark charge matrix is
\begin{equation}
\cQ = \diag \left( \frac{2}{3} , - \frac{1}{3} \right)
.\end{equation}
The operators contributing to baryon electromagnetic properties 
have the same form as in $SU(3)$. One must keep in mind, however, 
that the LECs appearing in Eqs.~\eqref{eq:local2}, \eqref{eq:local3}, and \eqref{eq:local5}
have different numerical values in $SU(2)$. These values are the same as
in $SU(4|2)$.  Additionally there are local electromagnetic operators 
in $SU(2)$ that involve the trace of the charge matrix. These
operators have the form
\begin{eqnarray}
\mathfrak{L} &=&
- 
\frac{e \ol c_T}{\L_\chi^2} 
\left( \ol T T \right)
v_\mu \partial_\nu F^{\mu \nu} \, \tr \cQ
- 
\frac{e \ol \mu_T \L_{QCD}}{4 \L_\chi m_Q} 
\left( \ol T \sigma_{\mu \nu} T \right) 
\, F^{\mu \nu} \, \tr \cQ 
\notag \\
&& + 
\frac{i e \ol \mu_S}{\L_\chi} 
\left( \ol S_\mu  S_\nu \right) \, F^{\mu \nu} \, \tr \cQ
+ 
\frac{e \ol c_S}{\L_\chi^2}
\left( \ol S {}^\a S_\a \right) \, v_\mu \partial_\nu F^{\mu \nu} \, \tr \cQ
- 
\frac{e \overline{\mathbb{Q}} {}_S }{ \L_\chi^2 } 
\left( \ol S {}^{\{ \mu} S^{\nu \}} \right) \, v_\a \partial_\mu F_{\nu}{}^{ \a} \, \tr \cQ
.\notag \\ \end{eqnarray}
The LECs appearing in $SU(4|2)$ are all numerically equal to those in the above Lagrangian for $SU(2)$.
Including contributions from these operators, the electromagnetic form factors of baryons in $SU(2)$ 
have exactly the same form as those in $SU(4|2)$ in Appendix~\ref{pqsutwo} with $q_{jl} = \frac{1}{3}$. 
For the $\L_Q$ baryon, however, the are no net loop contributions since $\a_\pi^{\L_Q} = 0$. 
For the $\Sigma_Q$ baryons, the value of the coefficient $\a$ is given by the isospin: $\a_\pi^{\Sigma_Q^{I}} = I$.

\section{Form factor relations in the heavy quark limit} \label{s:HQ}

In this Appendix, we derive relations between the electromagnetic form factors and transition form factors of the $B$ and $B^*$
baryons. These relations follow from the decoupling of the spins of heavy and light degrees of freedom 
in the heavy quark limit. We begin for simplicity with the case of the $s_\ell = 0$ baryons.

The electromagnetic current $J^\mu$ can be written in terms of light quark and heavy quark contributions, 
\begin{equation}
J^\mu = \sum_i \cQ_i \, \ol q_i \gamma^\mu q_i + \sum_i \cQ_{Q_i} \, \ol Q_i v^\mu Q_i
,\end{equation}
where the first sum is over the light flavors while the second is over the heavy flavors. 
Written this way, our arguments generalize to partially quenched theories. 
Notice that only the heavy quark charge operator survives the heavy quark limit, 
thus the action of $J^\mu$ between heavy quark states is proportional to unity.
Consider the current matrix elements of the $s_\ell = 0$ baryons. As the
spins of the heavy and light degrees of freedom are decoupled, the baryon states 
are direct products of $s_\ell = 0$ states and heavy quark states. The action of 
the current operator is always unity in the heavy quark subspace, thus we have
\begin{equation}
\langle \ol T (p') | J^\mu | T(p) \rangle  = \langle \ol Q | \mathbf{1} | Q \rangle \, v^\mu f(q^2)
.\end{equation}  
Comparing the above form with the current matrix element decomposition in Eq.~\eqref{eq:Tdecomp}, we 
find $F_1(q^2) \propto f(q^2)$ and $F_2(q^2) = 0$.

Now we consider the current matrix elements of the sextet baryons.
In the heavy quark limit, the $B$ and $B^*$ baryons are degenerate and are described by the $s_\ell = 1$ 
field $S^\mu$ defined in Eq.~\eqref{eq:Sdecomp}. Schematically current matrix elements of the $S$ field 
can be written as
\begin{equation}
\langle \ol S_\mu (p') | J^\rho | S_\nu(p) \rangle 
= 
\Big(  \langle s_\ell = 1, p' | \otimes \langle \ol Q | \Big) J^\rho  \Big(  | Q \rangle \otimes | s_\ell = 1, p \rangle \Big)
,\end{equation}
because of the decoupling of heavy and light spins. As the action of $J^\rho$ in the heavy 
quark sector is unity, current matrix elements of the $S$-field can be parametrized by
\begin{equation} \label{SJ}
\langle \ol S_\mu (p') | J^\rho | S_\nu(p) \rangle 
=
- \ol \varepsilon_\mu (p') \cO^{\mu \rho \nu} \varepsilon_\nu(p),
\end{equation}
where $\varepsilon_\mu(p)$ is the polarization spinor of the $S$-field that is proportional to
the direct product: $u(v) \otimes \epsilon_\mu(p)$, where $u(v)$ is the heavy quark spinor and
$\epsilon_\mu(p)$ is the polarization vector of the light degrees of freedom.
The polarization spinor satisfies the constraints 
$p \cdot \varepsilon(p) = \rlap \slash \varepsilon (p) = v \cdot \varepsilon (p) = 0$.
The tensor $\cO^{\mu \rho \nu}$ has the decomposition
\begin{equation}
\cO^{\mu \rho \nu} = g^{\mu \nu} v^\rho f_1(q^2) + \frac{1}{2 M_B} (q^\mu g^{\rho \nu} - q^\nu g^{\rho \mu}) f_2(q^2)
+ \frac{q^\mu q^\nu}{(2 M_B)^2} v^\rho g_1(q^2)
.\end{equation} 
Notice that there are only three form factors for the $S$-field because the light degrees of freedom have spin one. 
The above current matrix element encodes the electromagnetic form factors of both the $B$ 
and $B^*$ baryons, as well as their electromagnetic transition form factors.

Comparing the current matrix element of the $B^*$ fields in Eq.~\eqref{eq:B*decomp} with that of Eq.~\eqref{SJ}
allows us to determine
\begin{eqnarray}
F_1^*(q^2) &=& f_1(q^2), \notag \\
F_2^*(q^2) &=& f_2(q^2), \notag \\
G_1^*(q^2) &=& g_1(q^2), \text{ and} \notag \\
G_2^*(q^2) &=& 0,  
\end{eqnarray}
in the heavy quark limit. Notice that the magnetic octupole form factor $G_2^*(q^2)$ vanishes by 
heavy quark symmetry. Carrying out the procedure for the current matrix element of the $B$ fields 
in Eq.~\eqref{eq:Bdecomp}, we find
\begin{eqnarray}
F_1(q^2) &=& f_1(q^2) + \frac{q^2}{12 M_B^2} g_1(q^2), \text{ and} \notag \\
F_2(q^2) &=& \frac{2}{3} f_2(q^2),
\end{eqnarray}
in the heavy quark limit. Thus we have relations between the $B$ and $B^*$ electromagnetic form factors, namely
\begin{equation}
F_1(q^2) - F_1^*(q^2) = \frac{q^2}{12 M_B^2} G_1^*(q^2),
\end{equation}
and 
\begin{equation}
F_2(q^2) - \frac{2}{3} F_2^*(q^2) = 0 
.\end{equation}
These in turn yield relations between the static electromagnetic properties of the $B$ and $B^*$ baryons
\begin{equation}
<r^{*2}> - <r^2> = \frac{\mathbb{Q}}{M_B^2}
,\end{equation}
and
\begin{equation}
\mu = \frac{2}{3} \mu^*
.\end{equation}
These expressions are satisfied in the heavy quark limit by our one-loop \PQCPT\ and \CPT\ results.

Lastly we consider the electromagnetic transition matrix element
\begin{equation}
\langle \ol B {}^*_\mu (p') | J^\rho | B(p) \rangle 
= 
\ol u_\mu(p') \cO^{\mu \rho} \gamma_5 u(p)
,\end{equation}
where the tensor $\cO^{\mu \rho}$ is given by
\begin{equation}
\cO^{\mu \rho} 
= 
\left(g^{\mu \rho} - \frac{q^\mu \gamma^\rho}{2 M_B}\right) 
G_1(q^2)
- 
\frac{q^\mu v^\rho}{2 M_B} G_2(q^2)
+ 
\frac{1}{2 M_S^2} 
(q^2 g^{\mu \rho} - q^\mu q^\rho)
G_3(q^2)
.\end{equation}
Matching this decomposition onto that in Eq.~\eqref{SJ} for the $S$-field yields
\begin{eqnarray}
G_1(q^2) &=& \frac{1}{\sqrt{3}} f_2(q^2), \notag \\
G_2(q^2) &=& \frac{1}{\sqrt{3}} g_1(q^2), \text{ and} \notag \\
G_3(q^2) &=& 0
.\end{eqnarray}
In particular these relations imply that 
the magnetic dipole transition moment is given by
\begin{equation}
G_1(0) = \frac{1}{\sqrt{3}} \mu^*
,\end{equation}
while the electric quadrupole transition moment is given by
\begin{equation}
G_2(0) = - \frac{1}{2 \sqrt{3}} \mathbb{Q}
,\end{equation}
and the Coulomb quadrupole moment $G_3(0)$ vanishes.

\section{Quark Charges}\label{s:charge}

In this Appendix, we provide results for the more commonly used choice of the charge matrix $\cQ$. 
The choice used in the main text has advantages over this commonly used form as we explain below.
Insertion of different quark charges results only in modification of the charge dependent factors 
in our above results. Thus the only factors that are altered for a different choice of $\cQ$ are
the baryon charge $Q_T$ and $Q_B$, the charge of the light degrees of freedom $\cQ_T$ and $\cQ_S$,
and the coefficients $\a_\phi^T$ and $\a_\phi^B$ that depend on the charge of the loop meson $\phi$.

\begin{table}
\caption{The coefficients $\a^T_\phi$ in $SU(6|3)$ \PQCPT\ for the charge matrix $\cQ$ in Eq.~\eqref{eq:newQ}. 
Coefficients are listed for the $s_\ell = 0$ baryon states $T$, and are grouped into contributions from loop mesons
with mass $m_\phi$. The abbreviation $q_{jl}$ is defined to be $q_{jl} =  q_j + q_l$.}
\begin{tabular}{l | c c c c c c c }
    & $\qquad \pi \qquad \;$ & $\qquad K \qquad \;$ & $\qquad  \eta_s \qquad \;$ 
    & $ \qquad ju \qquad \;$ & $ \qquad ru \qquad \;$ 
    & $\qquad js \qquad \;$  & $\qquad rs \qquad \;$ \\
\hline
$\L_Q$     
	   &  $-\frac{1}{3} + q_{jl}$ & $\frac{1}{3} + q_r$  & $0$  
           &  $ \frac{1}{3} - q_{jl}$ & $\frac{1}{6} - q_r$  
           &  $0$ & $0$ \\
$\Xi^{+ \frac{1}{2}}_Q$     
	   &  $-\frac{1}{6} + \frac{1}{2} q_{jl}$ & $\frac{1}{2} (q_{jl} + q_r)$  & $\frac{1}{6} + \frac{1}{2} q_r$  
           &  $\frac{2}{3} - \frac{1}{2} q_{jl}$ & $\frac{1}{3} - \frac{1}{2} q_r$  
           &  $- \frac{1}{3} - \frac{1}{2} q_{jl}$ & $-\frac{1}{6} - \frac{1}{2} q_r$ \\
$\Xi^{- \frac{1}{2}}_Q$     
	   &  $- \frac{1}{6} + \frac{1}{2} q_{jl}$ & $\frac{1}{2} ( q_{jl} + q_r)$  & $\frac{1}{6} + \frac{1}{2} q_r$  
           &  $-\frac{1}{3} - \frac{1}{2} q_{jl}$ & $- \frac{1}{6} - \frac{1}{2} q_r $  
           &  $-\frac{1}{3} - \frac{1}{2} q_{jl}$ & $-\frac{1}{6} - \frac{1}{2} q_r$ \\
\end{tabular}
\label{t:newT}
\end{table}

\begin{table}
\caption{The coefficients $\a^B_\phi$ in $SU(6|3)$ \PQCPT\ for the charge matrix $\cQ$ in Eq.~\eqref{eq:newQ}. 
Coefficients are listed for the $s_\ell = 1$ baryon states $B$, and are grouped into contributions from loop mesons
with mass $m_\phi$. The abbreviation $q_{jl}$ is defined to be $q_{jl} =  q_j + q_l$.}
\begin{tabular}{l | c c c c c c c }
    & $\qquad \pi \qquad \;$ & $\qquad K \qquad \;$ & $\qquad  \eta_s \qquad \;$ 
    & $ \qquad ju \qquad \;$ & $ \qquad ru \qquad \;$ 
    & $\qquad js \qquad \;$  & $\qquad rs \qquad \;$ \\
\hline
$\Sigma^{+1}_Q$     
	   &  $-\frac{1}{3} + q_{jl}$ & $\frac{1}{3} + q_r$  & $0$  
           &  $ \frac{4}{3} - q_{jl}$ & $\frac{2}{3} - q_r$  
           &  $0$ & $0$ \\
$\Sigma^{0}_Q$     
	   &  $-\frac{1}{3} + q_{jl}$ & $\frac{1}{3} + q_r$  & $0$  
           &  $ \frac{1}{3} - q_{jl}$ & $\frac{1}{6} - q_r$  
           &  $0$ & $0$ \\
$\Sigma^{-1}_Q$     
	   &  $-\frac{1}{3} + q_{jl}$ & $\frac{1}{3} + q_r$  & $0$  
           &  $ -\frac{2}{3} - q_{jl}$ & $-\frac{1}{3} - q_r$  
           &  $0$ & $0$ \\
$\Xi^{\prime + \frac{1}{2}}_Q$     
	   &  $-\frac{1}{6} + \frac{1}{2} q_{jl}$ & $\frac{1}{2} (q_{jl} + q_r)$  & $\frac{1}{6} + \frac{1}{2} q_r$  
           &  $\frac{2}{3} - \frac{1}{2} q_{jl}$ & $\frac{1}{3} - \frac{1}{2} q_r$  
           &  $- \frac{1}{3} - \frac{1}{2} q_{jl}$ & $-\frac{1}{6} - \frac{1}{2} q_r$ \\
$\Xi^{\prime - \frac{1}{2}}_Q$     
	   &  $- \frac{1}{6} + \frac{1}{2} q_{jl}$ & $\frac{1}{2} ( q_{jl} + q_r)$  & $\frac{1}{6} + \frac{1}{2} q_r$  
           &  $-\frac{1}{3} - \frac{1}{2} q_{jl}$ & $- \frac{1}{6} - \frac{1}{2} q_r $  
           &  $-\frac{1}{3} - \frac{1}{2} q_{jl}$ & $-\frac{1}{6} - \frac{1}{2} q_r$ \\
$\Omega_Q$     
	   &  $0$ & $-\frac{1}{3} + q_{jl}$  & $\frac{1}{3} + q_r$  
           &  $0$ & $0$  
           &  $-\frac{2}{3} - q_{jl}$ & $-\frac{1}{3} - q_r$ \\
\end{tabular}
\label{t:newB}
\end{table}

The commonly used form of the charge matrix $\cQ$ in $SU(6|3)$ PQQCD is~\cite{Chen:2001yi}
\begin{equation} \label{eq:newQ}
\cQ = 
\diag 
\left( 
\frac{2}{3}, -\frac{1}{3}, -\frac{1}{3}, 
q_j, q_l, q_r, 
q_j, q_l, q_r
\right) 
.\end{equation}
When the sea and valence quark masses are made degenerate, one rather elegantly 
recovers QCD independent of the values of the charges $q_j$, $q_l$, and $q_r$. 
Implementation of this choice
for the charge matrix, however, is not the most economical in terms of computation time. 
This is because closed valence quark loops with a photon insertion are not
canceled by the corresponding ghost quark loops, as their charges are not identical. Thus the 
lattice practitioner must calculate closed valence quark loops with photon insertion (using
the effective charges $q_u = \frac{2}{3} - q_j$, $q_d = - \frac{1}{3} - q_l$, 
and $q_s = - \frac{1}{3} - q_ r$ to mimic the partial cancellation from ghost quark loops)
in addition to closed sea quark loops and closed sea quark loops with photon insertion. Closed valence 
quark loops without photon insertion are of course canceled by closed ghost loops because their charges never enter.
As valence quarks now appear in the closed loops with an operator insertion, the benefits of using 
lighter valence masses than sea masses cannot be maximally obtained. In the foreseeable future, 
the choice of $\cQ$ above is not ideal for performing partially quenched simulations.

\begin{table}
\caption{The coefficients $\a^T_\phi$ for the $\Lambda_Q$ in $SU(4|2)$ \PQCPT\ for the charge matrix $\cQ$ in Eq.~\eqref{eq:newQ2}. 
Coefficients are grouped into contributions from loop mesons with mass $m_\phi$. 
The abbreviation $q_{jl}$ is defined to be $ q_j + q_l$.}
\begin{tabular}{l | c c c c c c c }
    & $\qquad \eta_u \qquad \;$ & $\qquad \pi \qquad \;$ & $\qquad  \eta_d \qquad \;$ 
    & $ \qquad ju \qquad \;$ & $ \qquad lu \qquad \;$ 
    & $\qquad jd \qquad \;$  & $\qquad ld \qquad \;$ \\
\hline
$\Lambda_Q$     
	   &  $-\frac{1}{3} + \frac{1}{2} q_{j}$ & $- \frac{1}{6} + \frac{1}{2} q_{jl}$  & $\frac{1}{6} + \frac{1}{2} q_l$  
           &  $ \frac{1}{3} - \frac{1}{2} q_{j}$ & $\frac{1}{3} - \frac{1}{2} q_l$  
           &  $ - \frac{1}{6} - \frac{1}{2} q_j$ & $-\frac{1}{6} - \frac{1}{2} q_l$ \\
\end{tabular}
\label{t:newT2}
\end{table}

\begin{table}
\caption{The coefficients $\a^B_\phi$ in $SU(4|2)$ \PQCPT\ for the charge matrix $\cQ$ in Eq.~\eqref{eq:newQ2}. 
Coefficients are listed for the $s_\ell = 1$ baryon states $B$, and are grouped into contributions from loop mesons
with mass $m_\phi$. The abbreviation $q_{jl}$ is defined to be $q_{jl}=  q_j + q_l$.}
\begin{tabular}{l | c c c c c c c }
    & $\qquad \eta_u \qquad \;$ & $\qquad \pi \qquad \;$ & $\qquad  \eta_d \qquad \;$ 
    & $ \qquad ju \qquad \;$ & $ \qquad lu \qquad \;$ 
    & $\qquad jd \qquad \;$  & $\qquad ld \qquad \;$ \\
\hline
$\Sigma^{+1}_Q$     
	   &  $-\frac{2}{3} + q_{j}$ & $\frac{1}{3} + q_l$  & $0$  
           &  $ \frac{2}{3} - q_{j}$ & $\frac{2}{3} - q_l$  
           &  $0$ & $0$ \\
$\Sigma^{0}_Q$     
	   &  $-\frac{1}{3} + \frac{1}{2} q_{j}$ & $- \frac{1}{6} + \frac{1}{2} q_{jl}$  & $\frac{1}{6} + \frac{1}{2} q_l$  
           &  $ \frac{1}{3} - \frac{1}{2} q_{j}$ & $\frac{1}{3} - \frac{1}{2} q_l$  
           &  $-\frac{1}{6} - \frac{1}{2} q_j$ & $-\frac{1}{6} - \frac{1}{2} q_l$ \\
$\Sigma^{-1}_Q$     
	   &  $0$ & $-\frac{2}{3} + q_j$  & $\frac{1}{3} + q_l$  
	   &  $0$ & $0$	           
           &  $ -\frac{1}{3} - q_{j}$ & $-\frac{1}{3} - q_l$  \\           
\end{tabular}
\label{t:newB2}
\end{table}

In order that the valence and sea sectors be separated as their names suggest, one must have the ghost 
charges equal to their valence counterparts. With this choice, closed valence quark loops with operator insertion
are always canceled by the corresponding ghost loop. Thus the only quark disconnected contributions arise from 
the sea sector and these quarks efficaciously can be given larges masses to reduce computation time.
Thus a more practical choice of the $SU(6|3)$ charge matrix is that employed in the main text
\begin{equation} 
\cQ = \diag ( q_u, q_d, q_s, q_j, q_l , q_r, q_u, q_d, q_s ), \notag
\end{equation}
where to maintain supertracelessness $q_j + q_l + q_r = 0$. 
Notice that when the sea quark masses are made degenerate with 
the valence quark masses, QCD is only recovered for the specific choice 
$q_u = q_j = \frac{2}{3}$, and $q_d = q_s = q_l = q_r + - \frac{1}{3}$. 
One can use unphysical charges for both valence and sea quarks to determine the electromagnetic LECs.

To contrast with the calculation in $SU(6|3)$ in the main text, we calculate the baryon electromagnetic 
properties using the charge matrix in Eq.~\eqref{eq:newQ}.  The results have the same functional form 
as in the main text. One must be careful, however, to use the appropriate values of the baryon charge, the
charge of the light degrees of freedom, and replace the $SU(6|3)$ coefficients $\a_\phi^T$ and $\a_\phi^B$ with those
listed in Tables~\ref{t:newT} and \ref{t:newB}. Incomplete cancellations between valence and ghost sectors imply 
that valence-valence meson loops are not completely canceled by valence-ghost meson loops. 
Not surprisingly then these results have more complicated chiral extrapolation formula because more loop mesons contribute
and one can try to simplify the formula by using clever choices for the unfixed charges.

As with the case of $SU(6|3)$, the choice of charge matrix in~\cite{Beane:2002vq} for $SU(4|2)$ 
\begin{equation} \label{eq:newQ2}
\cQ = \diag \left( \frac{2}{3}, - \frac{1}{3}, q_j, q_l, q_j , q_l \right)
\end{equation}
is computationally intensive. A choice that more readily can be put to use on current lattices is
\begin{equation} \notag
\cQ = \diag (q_u, q_d, q_j, q_l, q_u, q_d )
,\end{equation}
which we used in Appendix~\ref{pqsutwo}. This choice mandates that the only closed quark loops are those with sea quarks.
Carrying out the calculation using $\cQ$ in Eq.~\eqref{eq:newQ2}, we find 
the modified $SU(4|2)$ coefficients $\a_\phi^T$ and $\a_\phi^B$ 
listed in Tables~\ref{t:newT2} and \ref{t:newB2}, respectively. 
Again there are more contributing loop mesons due to incomplete cancellations between the valence and ghost sectors of the theory.

\bibliography{hb}

\begin{thebibliography}{53}
\expandafter\ifx\csname natexlab\endcsname\relax\def\natexlab#1{#1}\fi
\expandafter\ifx\csname bibnamefont\endcsname\relax
  \def\bibnamefont#1{#1}\fi
\expandafter\ifx\csname bibfnamefont\endcsname\relax
  \def\bibfnamefont#1{#1}\fi
\expandafter\ifx\csname citenamefont\endcsname\relax
  \def\citenamefont#1{#1}\fi
\expandafter\ifx\csname url\endcsname\relax
  \def\url#1{\texttt{#1}}\fi
\expandafter\ifx\csname urlprefix\endcsname\relax\def\urlprefix{URL }\fi
\providecommand{\bibinfo}[2]{#2}
\providecommand{\eprint}[2][]{\url{#2}}

\bibitem[{\citenamefont{Alexandrou et~al.}(1994)}]{Alexandrou:1994dm}
\bibinfo{author}{\bibfnamefont{C.}~\bibnamefont{Alexandrou}}
  \bibnamefont{et~al.}, \bibinfo{journal}{Phys. Lett.}
  \textbf{\bibinfo{volume}{B337}}, \bibinfo{pages}{340} (\bibinfo{year}{1994}),
  \eprint{hep-lat/9407027}.

\bibitem[{\citenamefont{Bowler et~al.}(1996)}]{Bowler:1996ws}
\bibinfo{author}{\bibfnamefont{K.~C.} \bibnamefont{Bowler}}
  \bibnamefont{et~al.} (\bibinfo{collaboration}{UKQCD}),
  \bibinfo{journal}{Phys. Rev.} \textbf{\bibinfo{volume}{D54}},
  \bibinfo{pages}{3619} (\bibinfo{year}{1996}), \eprint{hep-lat/9601022}.

\bibitem[{\citenamefont{Ali~Khan et~al.}(2000)}]{AliKhan:1999yb}
\bibinfo{author}{\bibfnamefont{A.}~\bibnamefont{Ali~Khan}}
  \bibnamefont{et~al.}, \bibinfo{journal}{Phys. Rev.}
  \textbf{\bibinfo{volume}{D62}}, \bibinfo{pages}{054505}
  (\bibinfo{year}{2000}), \eprint{hep-lat/9912034}.

\bibitem[{\citenamefont{Woloshyn}(2000)}]{Woloshyn:2000fe}
\bibinfo{author}{\bibfnamefont{R.~M.} \bibnamefont{Woloshyn}},
  \bibinfo{journal}{Phys. Lett.} \textbf{\bibinfo{volume}{B476}},
  \bibinfo{pages}{309} (\bibinfo{year}{2000}), \eprint{hep-ph/0002088}.

\bibitem[{\citenamefont{Lewis et~al.}(2001)\citenamefont{Lewis, Mathur, and
  Woloshyn}}]{Lewis:2001iz}
\bibinfo{author}{\bibfnamefont{R.}~\bibnamefont{Lewis}},
  \bibinfo{author}{\bibfnamefont{N.}~\bibnamefont{Mathur}}, \bibnamefont{and}
  \bibinfo{author}{\bibfnamefont{R.~M.} \bibnamefont{Woloshyn}},
  \bibinfo{journal}{Phys. Rev.} \textbf{\bibinfo{volume}{D64}},
  \bibinfo{pages}{094509} (\bibinfo{year}{2001}), \eprint{hep-ph/0107037}.

\bibitem[{\citenamefont{Mathur et~al.}(2002)\citenamefont{Mathur, Lewis, and
  Woloshyn}}]{Mathur:2002ce}
\bibinfo{author}{\bibfnamefont{N.}~\bibnamefont{Mathur}},
  \bibinfo{author}{\bibfnamefont{R.}~\bibnamefont{Lewis}}, \bibnamefont{and}
  \bibinfo{author}{\bibfnamefont{R.~M.} \bibnamefont{Woloshyn}},
  \bibinfo{journal}{Phys. Rev.} \textbf{\bibinfo{volume}{D66}},
  \bibinfo{pages}{014502} (\bibinfo{year}{2002}), \eprint{hep-ph/0203253}.

\bibitem[{\citenamefont{Bowler et~al.}(1998)}]{Bowler:1997ej}
\bibinfo{author}{\bibfnamefont{K.~C.} \bibnamefont{Bowler}}
  \bibnamefont{et~al.} (\bibinfo{collaboration}{UKQCD}),
  \bibinfo{journal}{Phys. Rev.} \textbf{\bibinfo{volume}{D57}},
  \bibinfo{pages}{6948} (\bibinfo{year}{1998}), \eprint{hep-lat/9709028}.

\bibitem[{\citenamefont{Gottlieb and Tamhankar}(2003)}]{Gottlieb:2003yb}
\bibinfo{author}{\bibfnamefont{S.~A.} \bibnamefont{Gottlieb}} \bibnamefont{and}
  \bibinfo{author}{\bibfnamefont{S.}~\bibnamefont{Tamhankar}},
  \bibinfo{journal}{Nucl. Phys. Proc. Suppl.} \textbf{\bibinfo{volume}{119}},
  \bibinfo{pages}{644} (\bibinfo{year}{2003}), \eprint{hep-lat/0301022}.

\bibitem[{\citenamefont{Toussaint and Davies}(2004)}]{Toussaint:2004cj}
\bibinfo{author}{\bibfnamefont{D.}~\bibnamefont{Toussaint}} \bibnamefont{and}
  \bibinfo{author}{\bibfnamefont{C.~T.~H.} \bibnamefont{Davies}}
  (\bibinfo{year}{2004}), \eprint{hep-lat/0409129}.

\bibitem[{\citenamefont{Gasser and Leutwyler}(1984)}]{Gasser:1983yg}
\bibinfo{author}{\bibfnamefont{J.}~\bibnamefont{Gasser}} \bibnamefont{and}
  \bibinfo{author}{\bibfnamefont{H.}~\bibnamefont{Leutwyler}},
  \bibinfo{journal}{Ann. Phys.} \textbf{\bibinfo{volume}{158}},
  \bibinfo{pages}{142} (\bibinfo{year}{1984}).

\bibitem[{\citenamefont{Gasser and Leutwyler}(1985)}]{Gasser:1985gg}
\bibinfo{author}{\bibfnamefont{J.}~\bibnamefont{Gasser}} \bibnamefont{and}
  \bibinfo{author}{\bibfnamefont{H.}~\bibnamefont{Leutwyler}},
  \bibinfo{journal}{Nucl. Phys.} \textbf{\bibinfo{volume}{B250}},
  \bibinfo{pages}{465} (\bibinfo{year}{1985}).

\bibitem[{\citenamefont{Burdman and Donoghue}(1992)}]{Burdman:1992gh}
\bibinfo{author}{\bibfnamefont{G.}~\bibnamefont{Burdman}} \bibnamefont{and}
  \bibinfo{author}{\bibfnamefont{J.~F.} \bibnamefont{Donoghue}},
  \bibinfo{journal}{Phys. Lett.} \textbf{\bibinfo{volume}{B280}},
  \bibinfo{pages}{287} (\bibinfo{year}{1992}).

\bibitem[{\citenamefont{Wise}(1992)}]{Wise:1992hn}
\bibinfo{author}{\bibfnamefont{M.~B.} \bibnamefont{Wise}},
  \bibinfo{journal}{Phys. Rev.} \textbf{\bibinfo{volume}{D45}},
  \bibinfo{pages}{2188} (\bibinfo{year}{1992}).

\bibitem[{\citenamefont{Yan et~al.}(1992)}]{Yan:1992gz}
\bibinfo{author}{\bibfnamefont{T.-M.} \bibnamefont{Yan}} \bibnamefont{et~al.},
  \bibinfo{journal}{Phys. Rev.} \textbf{\bibinfo{volume}{D46}},
  \bibinfo{pages}{1148} (\bibinfo{year}{1992}).

\bibitem[{\citenamefont{Morel}(1987)}]{Morel:1987xk}
\bibinfo{author}{\bibfnamefont{A.}~\bibnamefont{Morel}}, \bibinfo{journal}{J.
  Phys. (France)} \textbf{\bibinfo{volume}{48}}, \bibinfo{pages}{1111}
  (\bibinfo{year}{1987}).

\bibitem[{\citenamefont{Sharpe}(1992)}]{Sharpe:1992ft}
\bibinfo{author}{\bibfnamefont{S.~R.} \bibnamefont{Sharpe}},
  \bibinfo{journal}{Phys. Rev.} \textbf{\bibinfo{volume}{D46}},
  \bibinfo{pages}{3146} (\bibinfo{year}{1992}),
  \eprint[http://arXiv.org/abs]{hep-lat/9205020}.

\bibitem[{\citenamefont{Bernard and Golterman}(1992)}]{Bernard:1992mk}
\bibinfo{author}{\bibfnamefont{C.~W.} \bibnamefont{Bernard}} \bibnamefont{and}
  \bibinfo{author}{\bibfnamefont{M.~F.~L.} \bibnamefont{Golterman}},
  \bibinfo{journal}{Phys. Rev.} \textbf{\bibinfo{volume}{D46}},
  \bibinfo{pages}{853} (\bibinfo{year}{1992}),
  \eprint[http://arXiv.org/abs]{hep-lat/9204007}.

\bibitem[{\citenamefont{Sharpe and Zhang}(1996)}]{Sharpe:1996qp}
\bibinfo{author}{\bibfnamefont{S.~R.} \bibnamefont{Sharpe}} \bibnamefont{and}
  \bibinfo{author}{\bibfnamefont{Y.}~\bibnamefont{Zhang}},
  \bibinfo{journal}{Phys. Rev.} \textbf{\bibinfo{volume}{D53}},
  \bibinfo{pages}{5125} (\bibinfo{year}{1996}),
  \eprint[http://arXiv.org/abs]{hep-lat/9510037}.

\bibitem[{\citenamefont{Labrenz and Sharpe}(1996)}]{Labrenz:1996jy}
\bibinfo{author}{\bibfnamefont{J.~N.} \bibnamefont{Labrenz}} \bibnamefont{and}
  \bibinfo{author}{\bibfnamefont{S.~R.} \bibnamefont{Sharpe}},
  \bibinfo{journal}{Phys. Rev.} \textbf{\bibinfo{volume}{D54}},
  \bibinfo{pages}{4595} (\bibinfo{year}{1996}),
  \eprint[http://arXiv.org/abs]{hep-lat/9605034}.

\bibitem[{\citenamefont{Bernard and Golterman}(1994)}]{Bernard:1994sv}
\bibinfo{author}{\bibfnamefont{C.~W.} \bibnamefont{Bernard}} \bibnamefont{and}
  \bibinfo{author}{\bibfnamefont{M.~F.~L.} \bibnamefont{Golterman}},
  \bibinfo{journal}{Phys. Rev.} \textbf{\bibinfo{volume}{D49}},
  \bibinfo{pages}{486} (\bibinfo{year}{1994}),
  \eprint[http://arXiv.org/abs]{hep-lat/9306005}.

\bibitem[{\citenamefont{Sharpe}(1997)}]{Sharpe:1997by}
\bibinfo{author}{\bibfnamefont{S.~R.} \bibnamefont{Sharpe}},
  \bibinfo{journal}{Phys. Rev.} \textbf{\bibinfo{volume}{D56}},
  \bibinfo{pages}{7052} (\bibinfo{year}{1997}),
  \eprint[http://arXiv.org/abs]{hep-lat/9707018}.

\bibitem[{\citenamefont{Golterman and Leung}(1998)}]{Golterman:1998st}
\bibinfo{author}{\bibfnamefont{M.~F.~L.} \bibnamefont{Golterman}}
  \bibnamefont{and} \bibinfo{author}{\bibfnamefont{K.-C.} \bibnamefont{Leung}},
  \bibinfo{journal}{Phys. Rev.} \textbf{\bibinfo{volume}{D57}},
  \bibinfo{pages}{5703} (\bibinfo{year}{1998}),
  \eprint[http://arXiv.org/abs]{hep-lat/9711033}.

\bibitem[{\citenamefont{Sharpe and Shoresh}(2000)}]{Sharpe:2000bc}
\bibinfo{author}{\bibfnamefont{S.~R.} \bibnamefont{Sharpe}} \bibnamefont{and}
  \bibinfo{author}{\bibfnamefont{N.}~\bibnamefont{Shoresh}},
  \bibinfo{journal}{Phys. Rev.} \textbf{\bibinfo{volume}{D62}},
  \bibinfo{pages}{094503} (\bibinfo{year}{2000}),
  \eprint[http://arXiv.org/abs]{hep-lat/0006017}.

\bibitem[{\citenamefont{Sharpe and Shoresh}(2001)}]{Sharpe:2001fh}
\bibinfo{author}{\bibfnamefont{S.~R.} \bibnamefont{Sharpe}} \bibnamefont{and}
  \bibinfo{author}{\bibfnamefont{N.}~\bibnamefont{Shoresh}},
  \bibinfo{journal}{Phys. Rev.} \textbf{\bibinfo{volume}{D64}},
  \bibinfo{pages}{114510} (\bibinfo{year}{2001}),
  \eprint[http://arXiv.org/abs]{hep-lat/0108003}.

\bibitem[{\citenamefont{Sharpe and Van~de Water}(2004)}]{Sharpe:2003vy}
\bibinfo{author}{\bibfnamefont{S.~R.} \bibnamefont{Sharpe}} \bibnamefont{and}
  \bibinfo{author}{\bibfnamefont{R.~S.} \bibnamefont{Van~de Water}},
  \bibinfo{journal}{Phys. Rev.} \textbf{\bibinfo{volume}{D69}},
  \bibinfo{pages}{054027} (\bibinfo{year}{2004}), \eprint{hep-lat/0310012}.

\bibitem[{\citenamefont{Chen and Savage}(2002{\natexlab{a}})}]{Chen:2001yi}
\bibinfo{author}{\bibfnamefont{J.-W.} \bibnamefont{Chen}} \bibnamefont{and}
  \bibinfo{author}{\bibfnamefont{M.~J.} \bibnamefont{Savage}},
  \bibinfo{journal}{Phys. Rev.} \textbf{\bibinfo{volume}{D65}},
  \bibinfo{pages}{094001} (\bibinfo{year}{2002}{\natexlab{a}}),
  \eprint[http://arXiv.org/abs]{hep-lat/0111050}.

\bibitem[{\citenamefont{Beane and Savage}(2002)}]{Beane:2002vq}
\bibinfo{author}{\bibfnamefont{S.~R.} \bibnamefont{Beane}} \bibnamefont{and}
  \bibinfo{author}{\bibfnamefont{M.~J.} \bibnamefont{Savage}},
  \bibinfo{journal}{Nucl. Phys.} \textbf{\bibinfo{volume}{A709}},
  \bibinfo{pages}{319} (\bibinfo{year}{2002}), \eprint{hep-lat/0203003}.

\bibitem[{\citenamefont{Chen and Savage}(2002{\natexlab{b}})}]{Chen:2002bz}
\bibinfo{author}{\bibfnamefont{J.-W.} \bibnamefont{Chen}} \bibnamefont{and}
  \bibinfo{author}{\bibfnamefont{M.~J.} \bibnamefont{Savage}},
  \bibinfo{journal}{Phys. Rev.} \textbf{\bibinfo{volume}{D66}},
  \bibinfo{pages}{074509} (\bibinfo{year}{2002}{\natexlab{b}}),
  \eprint{hep-lat/0207022}.

\bibitem[{\citenamefont{Arndt et~al.}(2003)\citenamefont{Arndt, Beane, and
  Savage}}]{Arndt:2003vx}
\bibinfo{author}{\bibfnamefont{D.}~\bibnamefont{Arndt}},
  \bibinfo{author}{\bibfnamefont{S.~R.} \bibnamefont{Beane}}, \bibnamefont{and}
  \bibinfo{author}{\bibfnamefont{M.~J.} \bibnamefont{Savage}},
  \bibinfo{journal}{Nucl. Phys.} \textbf{\bibinfo{volume}{A726}},
  \bibinfo{pages}{339} (\bibinfo{year}{2003}), \eprint{nucl-th/0304004}.

\bibitem[{\citenamefont{Arndt and Tiburzi}(2003{\natexlab{a}})}]{Arndt:2003ww}
\bibinfo{author}{\bibfnamefont{D.}~\bibnamefont{Arndt}} \bibnamefont{and}
  \bibinfo{author}{\bibfnamefont{B.~C.} \bibnamefont{Tiburzi}},
  \bibinfo{journal}{Phys. Rev.} \textbf{\bibinfo{volume}{D68}},
  \bibinfo{pages}{094501} (\bibinfo{year}{2003}{\natexlab{a}}),
  \eprint{hep-lat/0307003}.

\bibitem[{\citenamefont{Arndt and Tiburzi}(2003{\natexlab{b}})}]{Arndt:2003we}
\bibinfo{author}{\bibfnamefont{D.}~\bibnamefont{Arndt}} \bibnamefont{and}
  \bibinfo{author}{\bibfnamefont{B.~C.} \bibnamefont{Tiburzi}},
  \bibinfo{journal}{Phys. Rev.} \textbf{\bibinfo{volume}{D68}},
  \bibinfo{pages}{114503} (\bibinfo{year}{2003}{\natexlab{b}}),
  \eprint{hep-lat/0308001}.

\bibitem[{\citenamefont{Arndt and Tiburzi}(2004)}]{Arndt:2003vd}
\bibinfo{author}{\bibfnamefont{D.}~\bibnamefont{Arndt}} \bibnamefont{and}
  \bibinfo{author}{\bibfnamefont{B.~C.} \bibnamefont{Tiburzi}},
  \bibinfo{journal}{Phys. Rev.} \textbf{\bibinfo{volume}{D69}},
  \bibinfo{pages}{014501} (\bibinfo{year}{2004}), \eprint{hep-lat/0309013}.

\bibitem[{\citenamefont{Walker-Loud}(2004)}]{Walker-Loud:2004hf}
\bibinfo{author}{\bibfnamefont{A.}~\bibnamefont{Walker-Loud}}
  (\bibinfo{year}{2004}), \eprint{hep-lat/0405007}.

\bibitem[{\citenamefont{Tiburzi and Walker-Loud}(2004)}]{Tiburzi:2004rh}
\bibinfo{author}{\bibfnamefont{B.~C.} \bibnamefont{Tiburzi}} \bibnamefont{and}
  \bibinfo{author}{\bibfnamefont{A.}~\bibnamefont{Walker-Loud}}
  (\bibinfo{year}{2004}), \eprint{hep-lat/0407030}.

\bibitem[{\citenamefont{Balantekin and
  Bars}(1981{\natexlab{a}})}]{BahaBalantekin:1980pp}
\bibinfo{author}{\bibfnamefont{A.~B.} \bibnamefont{Balantekin}}
  \bibnamefont{and} \bibinfo{author}{\bibfnamefont{I.}~\bibnamefont{Bars}},
  \bibinfo{journal}{J. Math. Phys.} \textbf{\bibinfo{volume}{22}},
  \bibinfo{pages}{1810} (\bibinfo{year}{1981}{\natexlab{a}}).

\bibitem[{\citenamefont{Balantekin and
  Bars}(1981{\natexlab{b}})}]{BahaBalantekin:1981qy}
\bibinfo{author}{\bibfnamefont{A.~B.} \bibnamefont{Balantekin}}
  \bibnamefont{and} \bibinfo{author}{\bibfnamefont{I.}~\bibnamefont{Bars}},
  \bibinfo{journal}{J. Math. Phys.} \textbf{\bibinfo{volume}{22}},
  \bibinfo{pages}{1149} (\bibinfo{year}{1981}{\natexlab{b}}).

\bibitem[{\citenamefont{Golterman and Pallante}(2001)}]{Golterman:2001qj}
\bibinfo{author}{\bibfnamefont{M.}~\bibnamefont{Golterman}} \bibnamefont{and}
  \bibinfo{author}{\bibfnamefont{E.}~\bibnamefont{Pallante}},
  \bibinfo{journal}{JHEP} \textbf{\bibinfo{volume}{10}}, \bibinfo{pages}{037}
  (\bibinfo{year}{2001}), \eprint{hep-lat/0108010}.

\bibitem[{\citenamefont{Cho}(1992)}]{Cho:1992gg}
\bibinfo{author}{\bibfnamefont{P.~L.} \bibnamefont{Cho}},
  \bibinfo{journal}{Phys. Lett.} \textbf{\bibinfo{volume}{B285}},
  \bibinfo{pages}{145} (\bibinfo{year}{1992}), \eprint{hep-ph/9203225}.

\bibitem[{\citenamefont{Cho}(1993)}]{Cho:1992cf}
\bibinfo{author}{\bibfnamefont{P.~L.} \bibnamefont{Cho}},
  \bibinfo{journal}{Nucl. Phys.} \textbf{\bibinfo{volume}{B396}},
  \bibinfo{pages}{183} (\bibinfo{year}{1993}), \eprint{hep-ph/9208244}.

\bibitem[{\citenamefont{Savage}(1994)}]{Savage:1994zw}
\bibinfo{author}{\bibfnamefont{M.~J.} \bibnamefont{Savage}},
  \bibinfo{journal}{Phys. Lett.} \textbf{\bibinfo{volume}{B326}},
  \bibinfo{pages}{303} (\bibinfo{year}{1994}), \eprint{hep-ph/9401345}.

\bibitem[{\citenamefont{Banuls et~al.}(2000)\citenamefont{Banuls, Scimemi,
  Bernabeu, Gimenez, and Pich}}]{Banuls:1999mu}
\bibinfo{author}{\bibfnamefont{M.~C.} \bibnamefont{Banuls}},
  \bibinfo{author}{\bibfnamefont{I.}~\bibnamefont{Scimemi}},
  \bibinfo{author}{\bibfnamefont{J.}~\bibnamefont{Bernabeu}},
  \bibinfo{author}{\bibfnamefont{V.}~\bibnamefont{Gimenez}}, \bibnamefont{and}
  \bibinfo{author}{\bibfnamefont{A.}~\bibnamefont{Pich}},
  \bibinfo{journal}{Phys. Rev.} \textbf{\bibinfo{volume}{D61}},
  \bibinfo{pages}{074007} (\bibinfo{year}{2000}), \eprint{hep-ph/9905488}.

\bibitem[{\citenamefont{Chiladze}(1998)}]{Chiladze:1997uq}
\bibinfo{author}{\bibfnamefont{G.}~\bibnamefont{Chiladze}},
  \bibinfo{journal}{Phys. Rev.} \textbf{\bibinfo{volume}{D57}},
  \bibinfo{pages}{5586} (\bibinfo{year}{1998}), \eprint{hep-ph/9704426}.

\bibitem[{\citenamefont{Tiburzi}(2004)}]{Tiburzi:2004kd}
\bibinfo{author}{\bibfnamefont{B.~C.} \bibnamefont{Tiburzi}}
  (\bibinfo{year}{2004}), \eprint{hep-lat/0410033}.

\bibitem[{\citenamefont{Savage}(2002)}]{Savage:2001dy}
\bibinfo{author}{\bibfnamefont{M.~J.} \bibnamefont{Savage}},
  \bibinfo{journal}{Nucl. Phys.} \textbf{\bibinfo{volume}{A700}},
  \bibinfo{pages}{359} (\bibinfo{year}{2002}), \eprint{nucl-th/0107038}.

\bibitem[{\citenamefont{Hurni and Morel}(1983)}]{Hurni:1981ki}
\bibinfo{author}{\bibfnamefont{J.~P.} \bibnamefont{Hurni}} \bibnamefont{and}
  \bibinfo{author}{\bibfnamefont{B.}~\bibnamefont{Morel}}, \bibinfo{journal}{J.
  Math. Phys.} \textbf{\bibinfo{volume}{24}}, \bibinfo{pages}{157}
  (\bibinfo{year}{1983}).

\bibitem[{\citenamefont{Cho and Georgi}(1992)}]{Cho:1992nt}
\bibinfo{author}{\bibfnamefont{P.~L.} \bibnamefont{Cho}} \bibnamefont{and}
  \bibinfo{author}{\bibfnamefont{H.}~\bibnamefont{Georgi}},
  \bibinfo{journal}{Phys. Lett.} \textbf{\bibinfo{volume}{B296}},
  \bibinfo{pages}{408} (\bibinfo{year}{1992}), \eprint{hep-ph/9209239}.

\bibitem[{\citenamefont{Cheng et~al.}(1993)}]{Cheng:1992xi}
\bibinfo{author}{\bibfnamefont{H.-Y.} \bibnamefont{Cheng}}
  \bibnamefont{et~al.}, \bibinfo{journal}{Phys. Rev.}
  \textbf{\bibinfo{volume}{D47}}, \bibinfo{pages}{1030} (\bibinfo{year}{1993}),
  \eprint{hep-ph/9209262}.

\bibitem[{\citenamefont{Caswell and Lepage}(1986)}]{Caswell:1985ui}
\bibinfo{author}{\bibfnamefont{W.~E.} \bibnamefont{Caswell}} \bibnamefont{and}
  \bibinfo{author}{\bibfnamefont{G.~P.} \bibnamefont{Lepage}},
  \bibinfo{journal}{Phys. Lett.} \textbf{\bibinfo{volume}{B167}},
  \bibinfo{pages}{437} (\bibinfo{year}{1986}).

\bibitem[{\citenamefont{Manohar}(1997)}]{Manohar:1997qy}
\bibinfo{author}{\bibfnamefont{A.~V.} \bibnamefont{Manohar}},
  \bibinfo{journal}{Phys. Rev.} \textbf{\bibinfo{volume}{D56}},
  \bibinfo{pages}{230} (\bibinfo{year}{1997}), \eprint{hep-ph/9701294}.

\bibitem[{\citenamefont{Nozawa and Leinweber}(1990)}]{Nozawa:1990gt}
\bibinfo{author}{\bibfnamefont{S.}~\bibnamefont{Nozawa}} \bibnamefont{and}
  \bibinfo{author}{\bibfnamefont{D.~B.} \bibnamefont{Leinweber}},
  \bibinfo{journal}{Phys. Rev.} \textbf{\bibinfo{volume}{D42}},
  \bibinfo{pages}{3567} (\bibinfo{year}{1990}).

\bibitem[{\citenamefont{Jenkins and
  Manohar}(1991{\natexlab{a}})}]{Jenkins:1991jv}
\bibinfo{author}{\bibfnamefont{E.}~\bibnamefont{Jenkins}} \bibnamefont{and}
  \bibinfo{author}{\bibfnamefont{A.~V.} \bibnamefont{Manohar}},
  \bibinfo{journal}{Phys. Lett.} \textbf{\bibinfo{volume}{B255}},
  \bibinfo{pages}{558} (\bibinfo{year}{1991}{\natexlab{a}}).

\bibitem[{\citenamefont{Jenkins and
  Manohar}(1991{\natexlab{b}})}]{Jenkins:1991es}
\bibinfo{author}{\bibfnamefont{E.}~\bibnamefont{Jenkins}} \bibnamefont{and}
  \bibinfo{author}{\bibfnamefont{A.~V.} \bibnamefont{Manohar}},
  \bibinfo{journal}{Phys. Lett.} \textbf{\bibinfo{volume}{B259}},
  \bibinfo{pages}{353} (\bibinfo{year}{1991}{\natexlab{b}}).

\bibitem[{\citenamefont{Manohar and Wise}(2000)}]{Manohar:2000dt}
\bibinfo{author}{\bibfnamefont{A.~V.} \bibnamefont{Manohar}} \bibnamefont{and}
  \bibinfo{author}{\bibfnamefont{M.~B.} \bibnamefont{Wise}},
  \bibinfo{journal}{Cambridge Monogr. Part. Phys. Nucl. Phys. Cosmol.}
  \textbf{\bibinfo{volume}{10}}, \bibinfo{pages}{1} (\bibinfo{year}{2000}).

\end{thebibliography}
\end{document}